\documentclass[a4paper]{article}
\usepackage[dvips]{graphicx}
\usepackage{verbatim}
\usepackage{amssymb}
\usepackage{amsmath}
\usepackage{amsbsy}
\usepackage{amsfonts}
\usepackage{amsthm}
\usepackage{amsxtra}
\usepackage{t1enc}
\usepackage{bbm}
\usepackage{mathrsfs}
\usepackage{ifthen}
\input{epsf}

\begin{document}

\newtheoremstyle{thm-style-kalle}
{7pt}      
{7pt}      
{\itshape} 
{}         
{\scshape} 
{.}        
{.5em}     
{}         

\theoremstyle{thm-style-kalle}
    \newtheorem{theorem}{Theorem}[section]
    \newtheorem{proposition}[theorem]{Proposition}
    \newtheorem{corollary}[theorem]{Corollary}
    \newtheorem{lemma}[theorem]{Lemma}
    \newtheorem{definition}[theorem]{Definition}
    \newtheorem{example}{Example}[section]
    \newtheorem{remark}{Remark}[section]

\newenvironment{theoremn}[1] 
    {\begin{theorem} \emph{(#1)}}{\end{theorem}}
\newenvironment{theoremcn}[2] 
    {\begin{theorem} \cite{#1} \emph{(#2)}}{\end{theorem}}
\newenvironment{theoremc}[1] 
    {\begin{theorem} \cite{#1}}{\end{theorem}}
\newenvironment{propositionn}[1] 
    {\begin{proposition} \emph{(#1)}}{\end{proposition}}
\newenvironment{propositioncn}[2] 
    {\begin{proposition} \cite{#1} \emph{(#2)}}{\end{proposition}}
\newenvironment{definitionn}[1] 
    {\begin{definition} \emph{(#1)}}{\end{definition}}
\newenvironment{Proof}[1][Proof]{\begin{proof}[\sc{#1}]}{\end{proof}}

\newcommand{\Remark} {\emph{Remark:  }}
\newcommand{\QED} {$\quad \square$ \\}

\newcommand{\la} {\lambda}
\newcommand{\de} {\delta}
\newcommand{\be} {\beta}
\newcommand{\si} {\sigma}
\newcommand{\om} {\omega}
\newcommand{\te} {\theta}
\newcommand{\vphi} {\varphi}
\newcommand{\eps} {\varepsilon}
\newcommand{\Om} {\Omega}
\newcommand{\La} {\Lambda}
\newcommand{\Th} {\Theta}

\newcommand{\bH} {\mathbb{H}}  
\newcommand{\bC} {\mathbb{C}}  
\newcommand{\bR} {\mathbb{R}}  
\newcommand{\bN} {\mathbb{N}}  
\newcommand{\bZ} {\mathbb{Z}}  
\newcommand{\bQ} {\mathbb{Q}}  
\newcommand{\bD} {\mathbb{D}}  
\newcommand{\sD} {\mathcal{D}} %
\newcommand{\sH} {\mathcal{H}} %
\newcommand{\sL} {\mathcal{L}} %
\newcommand{\sA} {\mathcal{A}} %
\newcommand{\sV} {\mathcal{V}} %
\newcommand{\sP} {\mathcal{P}} %
\newcommand{\sS} {\mathcal{S}} %
\newcommand{\sR} {\mathcal{R}} %
\newcommand{\sF} {\mathcal{F}} %
\newcommand{\sM} {\mathcal{M}} %
\newcommand{\sT} {\mathcal{T}} %
\newcommand{\sK} {\mathcal{K}} %
\newcommand{\sU} {\mathcal{U}} %

\newcommand{\PR} {\mathsf{P}}
\newcommand{\EX} {\mathsf{E}}
\newcommand{\lE} {\Big\langle}
\newcommand{\rE} {\Big\rangle}
\newcommand{\qvl} {\langle}
\newcommand{\qvr} {\rangle}

\newcommand{\ud} {\mathrm{d}}
\newcommand{\bn} {\mathbf{n}}
\newcommand{\bm} {\mathbf{m}}
\newcommand{\bl} {\mathbf{l}}
\newcommand{\bg} {\mathbf{g}}
\newcommand{\bh} {\mathbf{h}}
\newcommand{\bv} {\mathbf{v}}
\newcommand{\bu} {\mathbf{u}}
\newcommand{\bfs} {\mathbf{f}}
\newcommand{\bs} {\mathbf{s}}
\newcommand{\bc} {\mathbf{c}}
\newcommand{\bte} {\boldsymbol{\theta}}
\newcommand{\bphi} {\boldsymbol{\phi}}
\newcommand{\bsi} {\boldsymbol{\sigma}}
\newcommand{\bta} {\boldsymbol{\tau}}
\newcommand{\bet} {\boldsymbol{\eta}}
\newcommand{\bel} {\boldsymbol{\ell}}
\newcommand{\bro} {\boldsymbol{\rho}}
\newcommand{\bde} {\boldsymbol{\delta}}
\newcommand{\bzero} {\boldsymbol{0}}
\newcommand{\bone} {\boldsymbol{1}}

\newcommand{\conn} {\leftrightarrow}
\newcommand{\nconn} {\nleftrightarrow}

\newcommand{\pder}[1] {\frac{\partial}{\partial #1}}
\newcommand{\ppder}[1] {\frac{\partial^2}{\partial #1^2}}
\newcommand{\Order} {\mathcal{O}}
\newcommand{\trans} {^{\mathsf{T}}}
\newcommand{\tp} {^{\mathsf{T}}}
\newcommand{\til} {\Tilde}
\newcommand{\wtil} {\widetilde}
\newcommand{\lrot} {\nabla \times}
\newcommand{\lgrad} {\nabla}
\newcommand{\ldiv} {\nabla \cdot}
\newcommand{\lapl} {\triangle}
\newcommand{\lconj} {\overline \nabla}
\newcommand{\unit} {\mathbf{1}}
\newcommand{\Imag} {\mathrm{Im } \; }
\newcommand{\Kern} {\mathrm{Ker } \; }
\newcommand{\End} {\mathrm{End } \; }
\newcommand{\Res}[1] {\mathrm{Res}_{#1} \; }
\newcommand{\oires}[2] {\frac{1}{2 \pi i}
    \oint_\infty \left( {#2} \right) \, \ud {#1} }
\newcommand{\oiresb}[2] {\frac{1}{2 \pi i}
    \oint_\infty \big( {#2} \big) \, \ud {#1} }
\newcommand{\oiresB}[2] {\frac{1}{2 \pi i}
    \oint_\infty \Big( {#2} \Big) \, \ud {#1} }
\newcommand{\oiresn}[2] {\frac{1}{2 \pi i}
    \oint_\infty {#2} \; \ud {#1} }
\newcommand{\cconj} {\overline}
\newcommand{\bbar} {\underline}
\newcommand{\tbar} {\overline}
\newcommand{\id} {\mathrm{id}}
\newcommand{\sign} {\mathrm{sign }}
\newcommand{\ann} {\mathbf{a}}
\newcommand{\cre} {\mathbf{a}^\dag}
\newcommand{\num} {\mathbf{N}}
\newcommand{\no} {\mathbf{:}}
\newcommand{\diam} {\mathrm{ diam}}
\newcommand{\conv} {*}
\newcommand{\ldual} {\big<}
\newcommand{\rdual} {\big>}
\newcommand{\bra} {\big<}
\newcommand{\ket} {\big>}
\newcommand{\im} {\Im \textrm{m }}
\newcommand{\re} {\Re \textrm{e }}

\newcommand{\cl} {\overline}
\newcommand{\bdr} {\partial}

\newcommand{\ag} {\hat{\mathfrak{g}}}
\newcommand{\ssg} {\mathfrak{g}}
\newcommand{\vir} {\mathfrak{vir}}
\newcommand{\sn} {\mathfrak{n}}
\newcommand{\irhwm} {\mathcal{H}}
\newcommand{\Verma} {\mathcal{V}}
\newcommand{\vac} {\Psi}
\newcommand{\hwvV} {\eta}
\newcommand{\hwvI} {\omega}
\newcommand{\zero} {\omega}
\newcommand{\ind} {\mathbf{1}}
\newcommand{\isom} {\cong}

\newcommand{\half} {\frac{1}{2}}
\newcommand{\const} {\mathrm{const.}}
\newcommand{\expand}[2] {|#1| \; \mathrm{ #2}}
\newcommand{\Expand}[2] {|#1| > |#2|}

\title{Virasoro Module Structure of
Local Martingales of SLE Variants}
\author{}
\date{}
\maketitle

\centerline{Kalle Kytölä}
\centerline{\small \texttt{kalle.kytola@cea.fr}}

\bigskip

\centerline{Service de Physique Th\'eorique de Saclay, CEA/DSM/SPhT}
\centerline{CEA-Saclay, 91191 Gif-sur-Yvette, France}

\bigskip

\begin{abstract}
Martingales often play an important role in computations with
Schramm-Loewner evolutions (SLEs).
The purpose of this article is to provide a straightforward
approach to the Virasoro module structure
of the space of local martingales for variants of SLEs.
In the case of ordinary chordal SLE, it has been shown in
Bauer\&Bernard: Phys.Lett.B {\bf 557}
that polynomial local martingales form a Virasoro module.
We will show for more general variants that the module of local
martingales has a natural submodule $\sM$ that has the same
interpretation as the module of polynomial local martingales
of chordal SLE, but it is in many cases easy to find more
local martingales than that.
We discuss the surprisingly rich structure of the
Virasoro module $\sM$ and construction of the ``SLE state'' or
``martingale generating function'' by Coulomb gas formalism.
In addition, Coulomb gas or Feigin-Fuchs integrals will be shown
to transparently produce candidates for multiple SLE pure
geometries.
\end{abstract}

\section{Introduction}
In \cite{Schramm-LERW_and_UST} Oded Schramm introduced the SLE
(stochastic Loewner evolution or Schramm-Loewner evolution) to describe
random conformally invariant curves by Loewner slit mapping technique.
The study of such objects is motivated by two dimensional
statistical mechanics at criticality. Continuum limits
of critical models, when they can be defined, are scale invariant and
it seems natural to expect conformal invariance as well.
SLE would then describe the continuum limits of curves or interfaces
in such models.
The introduction of SLE marked a leap in understanding geometric
questions in critical statistical mechanics.
However, the original definition of SLEs is quite restrictive what comes
to the boundary conditions it allows. To treat more general boundary
conditions, one uses variants of SLEs. Already the first papers
\cite{LSW-intersection_exponents_1, LSW-intersection_exponents_2,
LSW-intersection_exponents_3, RS-basic_properties, LSW-LERW_and_UST}
involved a couple of
variants, and later further generalizations have been explored.
This paper treats variants of quite general kind: we allow several
curves
and dependency on other marked points.

The question of continuum limit of critical models of statistical
mechanics has been studied by means of conformal field
theory (CFT) as well.
Roughly speaking CFT classifies local operators
by their transformation properties under local conformal
trasformations. Such classification resorts to representations of
Virasoro algebra.
The relation between SLEs and CFTs has attracted quite a lot of
attention recently, see e.g.
\cite{BB-CFTs_of_SLEs, FK-CFT_and_SLE, FW-restriction_representations,
Cardy-SLE_kappa_rho, KS-Malliavin_measures_SLE_and_CFT}.
CFT and a related method known as Coulomb gas have lead to numerous
succesful exact predictions about two dimensional models at criticality
during the past two and a half decades.
Applying the Coulomb gas approach to SLEs has also been considered in
the literature
\cite{Cardy-SLE_kappa_rho,
Kytola-SLE_kappa_rho, MRR-SLE_and_Coulomb_gas,
Gruzberg-stochastic_geometry}.

In this paper we will show that the space of local martingales for SLE
variants carries a representation of the Virasoro algebra --- thus
bringing the classification by conformal symmetry to natural SLE
quantities also. A group theoretic point of view behind this kind of
result was presented for the particular case of chordal SLE
in \cite{BB-SLE_martingales}.
The approach of this paper is more straightforward and concepts
that are needed are simpler (maybe at the loss of some elegance).
Furthermore, we will address the question of the structure of this
representation. It is remarkable that already when considering some
of the simplest SLE variants, many different kinds of representations of
the Virasoro algebra appear naturally: from irreducible
highest weight modules to quotients of Verma modules by
nonmaximal submodules and Fock spaces.

The Coulomb gas method will be studied as means of constructing
the ``SLE state'' (or martingale generating function). It will also
lead to explicit solutions of a system of differential equations
that are needed to define multiple SLEs
\cite{Dubedat-commutation,
BBK-multiple_SLEs, Graham-multiple_SLEs}
in a way much reminiscent of \cite{Dubedat-Euler_integrals}.
In \cite{BBK-multiple_SLEs} a conjecture about topological
configurations of multiple SLEs was presented.
Our explicit solutions are argued to be the ``pure geometries''
meant by that conjecture, that is multiple SLEs with a deterministic
topological configuration.

The paper is organized as follows.
In Section \ref{sec: SLE introduction} we introduce SLE
and give the definition appropriate for the purposes of this paper.
Section \ref{sec: BB philosophy} is an informal review of the idea of 
``SLE state'' (in the spirit of Bauer and Bernard), which
constitutes the core philosophy and heuristics underlying our results.
The main results of algebraic nature are then stated in Section
\ref{sec: Virasoro structure}:
we define a representation of 
the Virasoro algebra in a space of functions of
SLE data and show that local martingales form a
subrepresentation. A further natural submodule $\sM$ can be constructed
using nothing but the defining auxiliary function of the SLE variant
in question.
In the light of a few examples we make the first remarks about the
structure of this Virasoro module.
Section \ref{sec: Coulomb gas} briefly reviews some algebraic
aspects of the Coulomb gas method which are then applied
to constructions of SLE${_\kappa} (\underline{\rho})$ and
multiple SLE states. In particular concrete solutions to
the system of differential equations needed for multiple
SLE definition are obtained as Feigin-Fuchs integrals.
Finally in Section \ref{sec: discussion}, we digress to discuss
various aspects of the topics of earlier sections.
Choices of integration contours of screening charges are
argued to give rise to the ``pure geometries'',
we comment on fully M\"obius invariant SLE variants and
discuss prospects of completely resolving the structure of the
Virasoro module $\sM$ by BRST cohomology.

The general purpose of this paper is to exhibit a useful algebraic
structure of local martingales for the multiple SLEs, and to provide
a language and an elementary approach to this structure. The
approach has applications to SLE questions of different kinds,
in particular to the well known conjectures of chordal SLE
reversibility and SLE duality
\cite{KK-reversibility_duality, KKM-in_preparation}.


\section{Schramm-Loewner Evolutions (SLEs)}
\label{sec: SLE introduction}

\subsection{Curves in statistical mechanics at criticality and SLEs}


The realm of two dimensional models of statistical physics at their
critical point allows lots of exact results,
much owing to the observation that these models
often exhibit conformal invariance.
There is indeed a general argument that at criticality the continuum
limit of a two-dimensional model with local interactions is described
by a conformal field theory. Since 1980's, this approach to
studying the critical point has proved extremely powerful.
A key point in conformal field theory is to observe that we
can let Virasoro algebra act on local operators, thus vastly
reducing the amount of different operators needed to study.
The moral of this paper as well is that the action of Virasoro algebra
on an operator located at infinity allows us to build local martingales
for SLEs.
We will comment on this interpretation in Section
\ref{sec: role of M}.

While conformal field theory is traditional and successful,
Schramm's seminal article \cite{Schramm-LERW_and_UST} uses
another way of exploiting the presumed conformal invariance.
Instead of local objects the attention is directed to objects of
macroscopic scale. Whenever there exists a natural way of
defining an interface or curve of macroscopic size in the
lattice model, the same could be hoped for in its continuum
limit. The conformal invariance conjecture then concerns the
law of this curve in continuum limit.

To be more precise about the setup let us consider the case
that corresponds to chordal SLE, the simplest of SLE variants.
Imagine our model is defined in a simply connected
two dimensional domain $\Lambda \subset \bC$
and that there is a curve in the model
from point $a \in \bdr \Lambda$ to $b \in \bdr \Lambda$.
Let us denote by $\gamma_{\Lambda; a, b}$ the random curve
thus obtained.
The conformal invariance assumption states that for the
same model in another domain $\Lambda'$ such that there is
a curve from $a' \in \bdr \Lambda'$ to $b' \in \bdr \Lambda'$,
the law of $\gamma_{\Lambda'; a', b'}$ is the same as that of
the image of $\gamma_{\Lambda;a,b}$ under a conformal map
$f: \Lambda \rightarrow \Lambda'$ with $f(a)=a'$ and $f(b)=b'$.

In addition to the conformal invariance one needs another property
that is frequently satisfied by curves arising in models of
statistical mechanics.
If one considers the model conditioned on a piece of the curve
starting from $a$, say, then the result is often just the
same model in a subdomain with the piece of the curve
removed and the remaining part of the curve should now continue from
the tip of the removed piece.
This property is referred to as the domain Markov property.

It is an exquisite observation by Schramm that when one
uses Loewner's slit map technique to describe the curve
starting from $a$, then
the requirements of conformal invariance and domain Markov
property can be used together in a simple but powerful manner.
The conclusion is that there is a one parameter family of
probability measures on curves in $\overline{\Lambda}$
from $a$ to $b$ that satisfy the two requirements.
The sole significant parameter is called $\kappa \geq 0$.
For concreteness take
$\Lambda = \bH = \{ z \in \bC : \im z > 0 \}$, $a \in \bR$, $b=\infty$,
and $\gamma : [0, \infty) \rightarrow \overline{\bH}$ a continuous
parametrization of the random curve $\gamma_{\bH;a,\infty}$.
Then the conformal maps $g_t$ from the unbounded component of
$\bH \setminus \gamma[0,t]$ to $\bH$
satisfy $g_0(z)=z$ and the Loewner's equation
\begin{align*}
\ud g_t(z) \; = \; \frac{2}{g_t(z)-X_t} \; \ud \qvl A \qvr_t
\textrm{ ,}
\end{align*}
where $A_t$ is a continuous martingale,
$\qvl A \qvr_t$ its quadratic variation, and $X_0=a$,
$\ud X_t = \sqrt{\kappa} \; \ud A_t$.
The curve can be recovered through $\gamma(t) =
\lim_{\eps \downarrow 0} g_t^{-1}(X_t + i \eps)$,
see \cite{RS-basic_properties}.
The usual SLE terminology is the following:
$\gamma[0,\infty) \subset \overline{\bH}$ is called the SLE \emph{trace}
and by filling regions surrounded by the trace one obtains the
\emph{hull} $K_t$,
the closure of the complement of the unbounded component
of $\bH \setminus \gamma[0,t]$.
Thus $g_t : \bH \setminus K_t \rightarrow \bH$ is a
conformal map.

By now there are many very good
and comprehensive reviews of SLEs, e.g.
\cite{Werner-random_planar_curves, KN-guide, BB-2d_growth_processes,
Gruzberg-stochastic_geometry, Cardy-SLE_for_physicists},
each of them taking a different perspective to the topic.
In these the reader will find motivation, definitions,
history, properties and of course applications of SLE.

\subsection{Definition of SLE variants}
\label{sec: SLE definition}
The chordal SLE described above arises from simple boundary
conditions that ensure the existence of a curve from one boundary
point to another such that no other point plays a special role.
However, we may easily imagine our models with boundary conditions
that depend on other points and perhaps give rise to several curves.
We will thus give a less restrictive definition. However, to keep
the notation reasonable we allow these special points only at
the boundary. To allow marked points in the bulk, $z \in \Lambda$,
is a straightforward generalization (a bulk point can be treated
just as a pair of boundary points) but it would lead to an
unnecessarily heavy notation.

The definition is motivated by the connection of conformal field
theory and statistical mechanics, see e.g.
\cite{BBH-dipolar, BBK-multiple_SLEs}.
If the reader doesn't find this motivation sufficient, the use
of our definition can be justified by the fact that most SLE variants
proposed so far are covered by this definition:
chordal SLE${}_\kappa$, SLE${}_\kappa(\underline{\rho})$,
commonly used variants of multiple SLEs \cite{BBK-multiple_SLEs,
Graham-multiple_SLEs, Dubedat-Euler_integrals}, and with minor
changes radial SLE${}_\kappa$ and radial
SLE${}_\kappa(\underline{\rho})$ as well as mixed cases
\cite{SW-coordinate_changes}.

We will give the definition of SLE variants in the upper half plane
$\bH = \{ z \in \bC : \im z > 0 \}$. In other domains the SLEs are
defined by conformal invariance.

Let $\kappa>0$. 
There will be SLE curves starting
at points $X^{1}_0, \ldots, X^{N}_0 \in \bR = \bdr \bH$.
The curves look locally like chordal SLE${}_\kappa$ or chordal
SLE${}_{\kappa^*}$, $\kappa^* = 16/\kappa$. These are the
two values of kappa that can be consistently considered at the same
time \cite{Graham-multiple_SLEs, Dubedat-commutation} and the only
two that correspond to CFT of central charge
$c(\kappa) = \frac{1}{4} (6-\kappa)(6 - 16/\kappa)$.
Thus for $I=1,\ldots,N$ let $\kappa_I \in \{\kappa, 16/\kappa\}$.
We denote $h_{x_I} = \frac{6 - \kappa_I}{2 \kappa_I}$, in CFT
these are the conformal weights of boundary one-leg operators.

The ``boundary conditions'' may also depend on points
$Y^{1}_0, \ldots, Y^{M}_0 \in \bR$.
Numbers $h_{y_1}, \ldots, h_{y_M} \in \bR$ are 
parameters: in CFT they are the conformal weights of the
boundary 
(primary) operators at the points
$Y^{1}_0, \ldots, Y^{M}_0$.
The points $X^1_0, \ldots, X^N_0, Y^1_0, \ldots Y^M_0$
should be distinct. They will serve as initial conditions
for the stochastic processes $X^I_t$ and $Y^K_t$ defined below.

The definition of SLE variant consists of requirements for an
auxiliary function $Z$ (the partition function), system of stochastic
differential equations governing the driving processes $X^I_t$ and
passive points $Y^K_t$, and the
multiple slit Loewner equation for the uniformizing map $g_t$.
After listing these requirements we will
also recall the definitions of hull and traces, which are similar to
ordinary SLE definitions.


\bigskip

The auxiliary function $Z$ is a function of the
arguments $x_1, \ldots, x_N ; y_1, \ldots, y_M$ that are ordered on
the real line in the same way as
$X^1_0, \ldots, X^N_0 ; Y^1_0, \ldots, Y^M_0$. We assume the
following properties:
\begin{itemize}
\item[(a)] \emph{Smoothness and positivity:}
$Z$ is a smooth function of $x_1, \ldots, x_N, y_1, \ldots, y_M$
taking positive real values, that is
$Z \in C^\infty(S \rightarrow \bR_+)$, where $S \subset \bR^{N+M}$
is the set where the arguments
$x_1, \ldots, x_N, y_1, \ldots, y_M$
are ordered in the same way as
$X^1_0, \ldots, X^N_0, Y^1_0, \ldots, Y^M_0$.
\item[(b)] \emph{Null field equations:}
$Z$ is annihilated by the differential operators
\begin{align*}
\sD_I \; = \; & \frac{\kappa_I}{2} \; \frac{\partial^2}{\partial x_I^2} +
    \sum_{J \neq I} \big( \frac{2}{x_J - x_I} \; \pder{x_J}
    + \frac{(\kappa_J-6)/\kappa_J}{(x_J-x_I)^2} \big) \\
& + \sum_{K=1}^M \big( \frac{2}{y_K - x_I} \pder{y_K}
    - \frac{2 h_{y_K}}{(y_K-x_I)^2} \big)
\end{align*}
for all $I=1,\ldots,N$.
\item[(c)] \emph{Translation invariance:}
$Z(x_1+\sigma, \ldots, y_M+\sigma) = Z(x_1, \ldots, y_M)$
for all $\sigma \in \bR$.
\item[(d)] \emph{Homogeneity:}
For some $\Delta \in \bR$ and all $\lambda>0$ we have
$Z(\lambda x_1, \ldots, \lambda y_M) = \lambda^\Delta
Z(x_1, \ldots, y_M)$.
\end{itemize}
Sometimes we use only some of the assumptions or modifications of these.
We will try to make it explicit which properties are used at each step.
%

\bigskip
The ``driving processes'' $X^{I}_t$, $I=1,\ldots,N$, and
``passive points'' $Y^K_t$, $K=1,\ldots,M$, are assumed to
solve the system of It\^o differential equations
\begin{align}
\label{eq: driving processes}
\left\{
\begin{array}{ll}
\ud X^{I}_t \quad = \quad & \sqrt{\kappa_I} \; \ud A^{I}_t
    + \sum_{J \neq I} \frac{2}{X^{I}_t - X^{J}_t}
    \; \ud \qvl A^J \qvr_t \\
& \quad + \kappa_I \big( \partial_{x_I} \log Z)
    (X^{1}_t, \ldots, Y^{M}_t) \big) \; \ud \qvl A^I \qvr_t \\
\ud Y^K_t \quad = \quad & \sum_J \frac{2}{Y^K_t-X^J_t} \; \ud \qvl A^J \qvr_t
\end{array}
\right.
\textrm{ ,}
\end{align}
where the $A^I$ are continuous martingales, $\qvl A^I \qvr_t$ their
quadratic variations and the cross variations vanish,
$\qvl A^I , A^J \qvr_t = 0$ for $I \neq J$.
The solution is defined on a random time interval $t \in [0, \tau)$,
$\tau$ being for example the stopping time at which some of the
processes $X^1, \ldots, Y^M$ hit each other for the first time
or any stopping time smaller than that\footnote{It is sometimes
possible to continue the definition of an SLE consistently beyond the
first hitting time of the processes $X^I_t$ and $Y^K_t$,
by another SLE variant. The question is interesting and
frequently important, but for the purpose of this paper it has
little significance. However, in \cite{KK-reversibility_duality} the
interested reader can find an example application of the ideas of this
paper to a conjectural formulation of SLE duality that requires
consistent gluing of different SLE variants.}.
If $\frac{\ud \qvl A^I \qvr_t}{\ud t}$ exists, it's interpretation is
is the growth speed in terms
of half plane capacity of the $I^\textrm{th}$ curve at time $t$.

\bigskip

The growth process itself is encoded in a family of conformal
mappings $(g_t)_{t \in [0,\tau)}$, 
which are hydrodynamically normalized at infinity
\begin{align}
\label{eq: g coefs}
g_t(z) \; = \; z + \sum_{m \leq -2} g_m^{(t)} z^{1+m}
    \; = \; z + \Order (z^{-1})
\textrm{ .}
\end{align}
The conformal mappings are obtained from the Loewner equation
\begin{align}
\label{eq: Loewner}
\ud g_t(z) = \sum_{I=1}^N
    \frac{2}{g_t(z) - X^{I}_t} \; \ud \qvl A^I \qvr_t
\textrm{ ,}
\end{align}
with initial condition $g_0(z)=z$ for all $z \in \bH$.
The set $K_t$ is the closure in $\cl{\bH}$ of the complement
of the maximal set in which the solution of (\ref{eq: Loewner})
exists up to time $t$.
We call $K_t$ the hull of the SLE at time $t$ ---
it is compact, its complement $\bH \setminus K_t$ is simply connected
and $g_t : \bH \setminus K_t \rightarrow \bH$ is the unique conformal
map from $\bH \setminus K_t$ to $\bH$ with hydrodynamic normalization
(\ref{eq: g coefs}).

One defines the traces by $\gamma^I_t =
\lim_{\eps \downarrow 0} g_t^{-1}(X^I_t + i \eps)$. By absolute
continuity with respect to independent SLEs
one argues that the traces have the same almost sure properties as
ordinary SLE traces, see \cite{Graham-multiple_SLEs}. 
If $\kappa_I \leq 4$ the
trace $\gamma^I : [0, \tau) \rightarrow \cl{\bH}$ is a simple curve.
On the other hand, if $\kappa_I > 4$ the trace is a curve
with self intersections and if $\kappa_I \geq 8$ it is a space-filling
curve.
For example the fractal dimension of the trace $\gamma^I$ is almost
surely $\min \{ 1 + \frac{\kappa_I}{8} , 2 \}$ 
as shown in \cite{Beffara-dimension}.
Note also that for $\kappa \neq 4$ precisely one of the values
$\kappa, \kappa^*=16/\kappa$ corresponds to simple curves and one
to self-intersecting curves.

Although the traces of SLE are the random curves that
one is originally interested in, we hardly need them in this paper.
Rather, our purpose is to gain an algebraic insight to the
stochastic process defined by (\ref{eq: driving processes}) and
(\ref{eq: Loewner}), which will sometimes turn out useful for
studying the traces themselves.


\section{Prologue: SLE state \`a la Bauer \& Bernard}
\label{sec: BB philosophy}

\subsection{The state of the SLE quantum mechanics style}
Before 
even being precise about the
setup, let us comment on a general philosophy that allows to
guess how to build an appropriate Virasoro module of functions
of SLE data, which we will do in
Sections \ref{sec: the representation} and \ref{sec: local mgales}.
Here we intend to be impressionistic rather than precise, to
get an overall picture.
The idea resembles quantum mechanics: one wants to encode the state of
the SLE at each instant of time in a vector space. This vector
space carries a representation of the physical symmetries
of the problem --- in our case notably the conformal symmetry is
represented infinitesimally by Virasoro algebra.

The auxiliary function $Z$ has been argued to correspond to statistical
mechanics partition function of the underlying model with appropriate
boundary conditions \cite{BBH-dipolar, BBK-multiple_SLEs}.
In conformal field theory this should be a correlation function
of the (primary) fields implementing the boundary conditions
\begin{align*}
Z(x_1, \ldots, y_M) = \bra \psi_{\delta_\infty}(\infty) \;
    \prod_K \psi_{\delta_{y_K}}(y_K) \prod_I \psi_{\delta_{x_I}} (x_I)
    \ket^{CFT}_{\bH}
\textrm{ .}
\end{align*}
In the operator formalism of conformal field theory this is written as
\begin{align*}
Z(x_1, \ldots, y_M) = \bra \omega^*_\infty ,
    \Psi(x_1,\ldots,x_N;y_1,\ldots,y_M) \omega_0 \ket
\textrm{ ,}
\end{align*}
where $\omega_0$ is the absolute vacuum,
$\Psi$ is a ``composition'' of intertwining operators
and $\omega^*_\infty$ is a vacuum whose conformal weight
is that of the operator at infinity. 
\begin{remark}
The $\delta_{y_K}$ and $\delta_{x_I}$ are conformal weights of
the boundary primary fields and should be the same as
$h_{y_K}$ and $h_{x_I}$.
But for the moment let us keep them as free parameters, it is
instructive to see at which point we will have to fix their
values.
\end{remark}

To create the state of SLE, one should start from the absolute
vacuum $\omega_0$ of CFT in the half plane,
apply the operator $\Psi$, implement the conformal map $g_t^{-1}$ by
an operator $G_{g_t}$, and normalize by the partition function $Z$:
\begin{align*}
M_t = \frac{1}{Z(X^1_t, \ldots, Y^M_t)} \; G_{g_t} \;
    \Psi (X^1_t, \ldots, X^N_t; Y^1_t, \ldots, Y^M_t) \; \omega_0
\textrm{ .}
\end{align*}
While $Z$ corresponded to the partition function, the ratios
\begin{align}
\label{eq: correlation function}
\frac{\bra u^* , G_{g_t} \Psi(X^1_t, \ldots, Y^M_t) \omega_0 \ket}
    {Z(X^1_t, \ldots, Y^M_t)}
\; = \; \bra u^* , M_t \ket
\end{align}
for any dual vectors $u^*$ correspond to correlation functions conditioned
on information at time $t$, see
\cite{BBH-dipolar, BBK-multiple_SLEs}.
The state $M_t$ in the state space of the conformal field theory
would be a vector valued local martingale,
a kind of martingale generating function
\cite{BB-CFTs_of_SLEs, BB-SLE_martingales, BB-conformal_transformations}.

\subsection{The role of the Virasoro module}
\label{sec: role of M}
We expect the space that we are working in to carry a representation
of the Virasoro algebra.
We recall that the Virasoro algebra $\vir$ is the Lie algebra spanned
by $L_n$, $n \in \bZ$, and $C$ with the commutation relations
\begin{align*}
[L_n, L_m] = (n-m) \; L_{n+m} + \frac{1}{12} (n^3-n) \delta_{n+m,0} \; C
\quad \textrm{ and } \quad
[C, L_n] = 0 
\textrm{ .}
\end{align*}
The central element $C$
acts as a multiplication by a number $c \in \bC$ in all the
representations we will study.
This number is called the central charge.
If we need several representations simultaneously,
$c$ takes the same value in all of them.

In (\ref{eq: correlation function}) we are free to project
to any dual vector $u^*$. A trivial thing to do is to choose
$u^* = \omega^*_\infty$, in which case the numerator is
also $Z$ and the ratio (\ref{eq: correlation function}) is
constant $1$, obviously a (local) martingale.

But since the dual also carries a representation of $\vir$ defined by
$\bra L_n u^* , u \ket = \bra u^* , L_{-n} u \ket$, one easily
obtains more interesting correlation functions. We can choose
$u^* = L_{-n_1} \cdots L_{-n_k} \omega^*_\infty$ and
thus build a whole highest weight module
\begin{align*}
\bra \sU(\vir) \omega^*_\infty ,
    G_{g_t} \Psi(X^1_t, \ldots, Y^M_t) \omega_0 \ket
\end{align*}
of these. In the rest of the paper what is denoted by $\sM$ will
play the role of this module. Morally it appears as the
contravariant representation of the space in which the SLE
state is encoded.
Thus is should be interpreted as consisting of the descendants
of the local operator at infinity.

\subsection{Explicit form of the representation}
Above it was argued that there should exist a Virasoro module consisting
of local martingales. Let us now give a little concreteness to these
thoughts. We should take a closer look at a couple of objects that
appeared in the discussion: the vacua $\omega_0$ and
$\omega^*_\infty$, the operator $G_{g_t}$ implementing
conformal transformation $g^{-1}_t$ and the intertwining operator
$\Psi$.

The absolute vacuum $\omega_0$ in conformal field theory is a
highest weight state of weight $0$, in other words it is a
singular vector $L_n \omega_0 = 0$ for all $n>0$ and has the
$L_0$ eigenvalue $0$, $L_0 \omega_0 = 0$. Moreover the vacuum $\omega_0$
should be translation invariant and since $L_{-1}$ represents an
infinitesimal translation this means $L_{-1} \omega_0 = 0$. These
observations say that (generically) the $\vir$ module
generated by $\omega_0$ is an irreducible highest weight module
of highest weight $0$.

It is not as obvious that $\omega^*_\infty$ should be the
absolute vacuum. If it were, we would at least have
$\delta_\infty = 0$. This case is related to M\"obius invariance
and it deserves a separate discussion,
Section \ref{sec: Mobius invariance}. But for now we only assume
that $\omega^*_\infty$ is a singular vector,
$L_n \omega^*_\infty = 0$ for $n>0$, and has weight
$\delta_\infty$, $L_0 \omega^*_\infty =
\delta_\infty \, \omega^*_\infty$. These assumptions
mean that the boundary 
operator at infinity is primary.

In the operator formalism of CFT, to a primary field
$\psi_\delta(x)$ of conformal weight $\delta$ corresponds
an intertwining operator $\Psi_\delta(x)$ from one Virasoro
module to another. The intertwining relations
$[L_n, \Psi_\delta(x)] = (x^{1+n} \pder{x} + (1+n) \delta x^n)
\, \Psi_\delta(x)$ are the infinitesimal form of the
transformation property $\psi_\delta(x) \overset{f}{\longrightarrow}
f'(x)^\delta \psi_\delta(f(x))$ of the primary field
under conformal transformations $f$. Our operator
$\Psi (x_1, \ldots, x_N ; y_1, \ldots, y_M)$ should
be composed of several intertwining operators and thus it should
have the intertwining property
\begin{align*}
[L_n , \Psi(\cdots)] \; = \; & \Big(
    \sum_I \big( x_I^{1+n} \pder{x_I} + (1+n) \delta_{x_I} x_I^n \big) \\
& + \sum_K \big( y_K^{1+n} \pder{y_K} + (1+n) \delta_{y_K} y_K^n \big) \Big)
    \Psi(\cdots)
\end{align*}

Finally we discuss the operator $G_{f}$ implementing the inverse
of a hydrodynamically normalized conformal map $f$ whose power
series expansion at infinity is
$f(z) = z + \sum_{l \leq -2} f_l z^{1+l}$. The construction of
$G_f$ was done in \cite{BB-conformal_transformations}.
The operator takes values in the completion of the universal
enveloping algebra of negative generators of $\vir$, that is
$G_f \in \overline{\sU(\vir_-)}$, and the mapping $f \mapsto G_f$
is a group anti-homomorphism. The defining properties are
$G_{\id_\bH} = 1$ and
\begin{align}
\label{eq: defining G}
\pder{f_m} G_f = & -\sum_{k \leq m} \oires{w}{w^{1+m}
    \frac{f'(w)}{f(w)^{2+k}}} \; G_f L_k
\textrm{ .}
\end{align}
Conversely it was also computed that for $k \leq -2$ we have
\begin{align*}
G_f L_k = &
    - \sum_{l \leq -2} \oires{z}{z^{-2-l} f(z)^{1+k}} \; \pder{f_l} G_f
\textrm{ .}
\end{align*}
In addition Bauer and Bernard showed that under conjugation by
$G_f$, $L_n$ transforms in the following way
\begin{align*}
G_f^{-1} L_n G_f = & \frac{c}{12} \oires{u}{u^{1+n} Sf(u)}
    + \sum_{k \leq n} \oires{u}{u^{1+n} \frac{f'(u)^2}{f(u)^{2+k}}} \; L_k
\textrm{ ,}
\end{align*}
where $Sf(z) = \frac{f'''(z)}{f'(z)}
- \frac{3}{2} \big(\frac{f''(z)}{f'(z)} \big)^2$ is the Schwarzian
derivative.
This is the transformation formula of the modes of stress
tensor under the conformal map $f$.

The properties of $G_f$ above can be combined, by
separating $k \leq -2 $ and $-1 \leq k$ in $G_f^{-1} L_{-n} G_f$,
to yield
\begin{align*}
& G_f^{-1} L_{-n} G_f \\
= \; & \frac{c}{12} \oires{u}{u^{1-n} Sf(u)}
    + \sum_{-1 \leq k \leq -n} \oires{u}{u^{1-n}
    \frac{f'(u)^2}{f(u)^{2+k}}} \; L_k \\
& + \sum_{l \leq -2} \oires{u}{\oires{z}{u^{1-n} f'(u)^2 z^{-2-l}
    \frac{1}{f(z)-f(u)}}} \; \pder{f_l} G_f
\textrm{ .}
\end{align*}
Using the intertwining property of $\Psi$ to commute the $L_k$,
$k \geq -1$, to the right we obtain for all $u^*$ the formula
\begin{align*}
\bra L_n u^*, G_f \Psi (x_1,\ldots,y_M) \omega_0 \ket
\; = \; & \bra u^*, L_{-n} G_f \Psi (x_1,\ldots,y_M) \omega_0 \ket \\
\; = \; & \bra u^*, G_f (G_f^{-1} L_{-n} G_f)
     \Psi (x_1,\ldots,y_M) \omega_0 \ket \\
\; = \; & \sL_n \; \bra u^*, G_f \Psi(x_1,\ldots,y_M) \omega_0 \ket
\textrm{ ,}
\end{align*}
where $\sL_n$ is the differential operator
\begin{align}
\sL_n \; 
\nonumber
= \; & \frac{c}{12} \oires{u}{u^{1-n} Sf(u)}
    - \sum_{l \leq 2} \oiresn{u}{\oires{z}{\frac{u^{1-n} f'(u)^2 z^{-2-l}}
    {f(z) - f(u)}} } \; \pder{f_l} \\
\nonumber
& + \sum_{I} \Big\{ \oires{u}{\frac{\delta_{x_I}}{(f(u)-x_I)^2}} +
    \oires{u}{\frac{u^{1-n} f'(u)^2}{f(u)-x_I}} \; \pder{x_I} \Big\} \\
\label{eq: Ln naive formula}
& + \sum_{K} \Big\{ \oires{u}{\frac{\delta_{y_K}}{(f(u)-y_K)^2}} +
    \oires{u}{\frac{u^{1-n} f'(u)^2}{f(u)-y_K}} \; \pder{y_K} \Big\}
\textrm{ .}
\end{align}
Recalling that $Z = \bra \omega^*_\infty , \Psi(\cdots) \omega_0 \ket
= \bra \omega^*_\infty , G_f \Psi(\cdots) \omega_0 \ket$,
the Virasoro module $\sM$
can be constructed starting from $Z$ and
recursively applying the differential operators $\sL_{n}$ above.
Note that if $\omega^*_\infty$ is indeed a singular
vector and a weight vector, using only $n<0$ will be sufficient.

\subsection{The plan and remarks about earlier work}
If we have faith in the above philosophy we now have at least
two possible ways to proceed.
One would be to construct explicitly the state $M_t$ in an
appropriate space that carries a representation of
Virasoro algebra and check that it is indeed a vector valued local
martingale\footnote{This is the approach that was succesfully
applied to chordal SLE in
\cite{BB-CFTs_of_SLEs, BB-conformal_transformations}.}.
The other one is to more or less forget about the above discussion
and just check that the procedure of applying the explicitly given
operators $\sL_n$ allows us to build local martingales starting
from $Z$.

The advantage of the former way is obviously that it makes direct
contact with quantum field theory.
The state $M_t$ encodes the information of the SLE at time $t$ ---
all of it if we are lucky (or smart).
We will indeed take on the task of constructing $M_t$ for some
cases of particular interest in Section \ref{sec: Coulomb gas}.

The latter way might seem slightly brutal, especially considering
the not particularly elegant formula (\ref{eq: Ln naive formula}).
But the straightforwardness has its own advantages as we will
see. The concepts needed are certainly simpler: the work boils
down to studying the first order differential operators $\sL_n$
in relation with the SLE process. One need not know
anything but the definition of the SLE itself (the partition
function $Z$ being a part of it).
In particular, we can start to work without taking a stand on the
question in which space $M_t$ is supposed to live.
One might make a natural guess that a highest weight
module for $\vir$ is appropriate (maybe irreducible,
maybe the Verma module or maybe the quotient of Verma module
by a non maximal submodule),
but in fact it will turn out that this is not possible even
in some of the simplest cases --- remarkably a coordinate
transform of the chordal SLE.

We mainly follow the latter approach due to its virtue of
straightforwardness. Yet a small benefit is that it offers
something of an alternative to the formalism that has already
been presented in the literature
\cite{BB-SLE_martingales, BB-CFTs_of_SLEs,
BB-conformal_transformations}.
Only Section \ref{sec: Coulomb gas} will address the question of
constructing $M_t$.

\section{The Virasoro module of local martingales}
\label{sec: Virasoro structure}
In this section we state the main results about the representation
of Virasoro algebra in the space of local martingales.
To define the representation,
we need some preliminaries about formal distributions which are
provided in Section \ref{sec: formal distr}. Next we discuss
a concept of homogeneity in Section \ref{sec: homogeneity}.
After having defined the
representation on the space of functions of SLE data, Section
\ref{sec: the representation}, we will show,
in Section \ref{sec: local mgales}, that local martingales form
a subrepresentation. We also show that the very natural further
submodule $\sM$ is a highest weight representation if
the auxiliary function $Z$ is translation invariant and homogeneous.

\subsection{Functions of SLE data}
\label{sec: SLE data}
The information about the SLE state at time $t$ consists
of the hull $K_t \subset \overline{\bH}$ and positions at which
the special points are located, that is the tips of the $N$ traces
and $M$ marked points. The hull is alternatively encoded
in the uniformizing map $g_t : \bH \setminus K_t \rightarrow \bH$,
and $g_t$ takes the tips of the traces to $X^1_t, \ldots, X^N_t$
and marked points to $Y^1_t, \ldots, Y^M_t$. Furthermore, the
map $g_t$ is uniquely determined by its expansion at infinity
(\ref{eq: g coefs}). Therefore, the information
can be represented by the infinite list of real valued
stochastic processes
\begin{align*}
X^1_t, \ldots, X^N_t \; ; \; Y^1_t, \ldots, Y^M_t \; ; \;
g_{-2}^{(t)}, g_{-3}^{(t)}, g_{-4}^{(t)}, \ldots
\end{align*}
governed by equations (\ref{eq: driving processes}) and
(\ref{eq: Loewner}).

The kind of local martingales we want to build are functions
of $X^I_t$, $Y^K_t$ and $g_{l}^{(t)}$.
More precisely, we are looking for functions of variables
$x_1, \ldots, x_N$; $y_1, \ldots, y_M$; $f_{-2}, f_{-3}, \ldots$
such that for any such function $\eta$, the ratio
\begin{align*}
\frac{\eta(x_1, \ldots, x_N; y_1, \ldots, y_M; f_{-2}, f_{-3}, \ldots)}
{Z(x_1, \ldots, x_N)}
\end{align*}
evaluated at $x_I = X^I_t$, $y_K = Y^K_t$, $f_{l} = g_{l}^{(t)}$
is a local martingale.

The proposed operators $\sL_n$ in (\ref{eq: Ln naive formula})
contain infinitely many terms: for each $l$ there is a term
$(\cdots) \pder{f_l}$. In order to avoid convergence problems
we only consider polynomials in the variables
$f_{-2}, f_{-3}, \ldots$. We also need to differentiate in variables
$x_I$ and $y_K$. Therefore we choose to work with functions from
the space
\begin{align*}
\sF = C^\infty(S \rightarrow \bC) [f_{-2}, f_{-3}, \ldots]
\textrm{ ,}
\end{align*}
where $S$ is the subset of $\bR^{N+M}$ where the variables
$x_1, \ldots, x_N, y_1, \ldots, y_M$ are ordered in the same
way as the initial conditions
$X^1_0, \ldots, X^N_0, Y^1_0, \ldots, Y^M_0$.
\begin{remark}
The algebra is rather independent of our choice of
function spaces. One may want to replace $\sF$
by some other space and as long as it is possible to make sense
of the operators and establish Corollaries \ref{cor: Virasoro} and
\ref{cor: A}, this is no problem.
\end{remark}

\subsection{Formal distributions}
\label{sec: formal distr}
This section will briefly recall the basics of formal distributions
as they will soon be needed.
A good treatment of the subject can be found e.g. in
\cite{Kac-vertex_algebras} and we use some results whose proofs are
easiest found there.

For $U$ a vector space, we denote by $U[[z, z^{-1}, w, w^{-1}, \ldots]]$
the set of formal expressions of type
\begin{align*}
\sum_{m,n,\ldots \in \bZ} a_{m,n,\ldots} \; z^m w^n \cdots
\textrm{ ,}
\end{align*}
where $a_{m,n,\ldots} \in U$. We call expressions of this type
formal distributions in the indeterminates $z,w,\ldots$ with 
coefficients in $U$. Important subspaces include series with only
non-negative/non-positive powers, finite series and
semi-infinite series,
\begin{align*}
U[[z]] \; := \; & \big\{ \sum_{m \in \bN} a_m z^m \; : \;
    \textrm{$a_m \in U$ for all $m \in \bN$} \big\} \\
U[z,z^{-1}] \; := \; & \big\{ \sum_{m \in \bZ} a_m z^m \; : \;
    \textrm{$a_m \in U$ for all $m \in \bZ$ and $a_m = 0$ for
    $|m| \gg 0$} \big\} \\
U((z)) \; := \; & \big\{ \sum_{m \in \bZ} a_m z^m \; : \;
    \textrm{$a_m \in U$ for all $m \in \bZ$ and $a_m = 0$ for
    $m \ll 0$} \big\}
\textrm{ .}
\end{align*}
We use similar notation for several variables.

The residue of a formal distribution is defined by
\begin{align*}
\Res{z} \sum_{m \in \bZ} a_m z^m = a_{-1}
\textrm{ .}
\end{align*}
A formal distribution
$\sum_{m,n,\ldots \in \bZ} a_{m,n,\ldots} \; z^m w^n \cdots$
can also be viewed as a formal distribution in the indeterminate
$z$ with coefficients in $U[[w,w^{-1},\ldots]]$,
so we can understand $\Res{z} a(z,w,\ldots) \in U[[w,w^{-1},\ldots]]$.

In this paper all vector spaces are over $\bC$, and
$U$ is usually an associative algebra, $U = \bC$ or $U = \End V$.
Thus we have naturally defined products e.g.
$\bC [z, z^{-1}] \times U[[z,z^{-1}]] \rightarrow U[[z,z^{-1}]]$
and
$U ((z^{-1})) \times U((z^{-1})) \rightarrow U((z^{-1}))$.
Note that whenever $a(z) b(z)$ is defined, the Leibniz's rule
$\partial_z ( a(z) b(z) ) = (\partial_z a(z) ) b(z) + a(z) (\partial_z b(z))$
and $\Res{z} \partial_z c(z) = 0$ lead to an integration by parts formula.

We denote the two different power series expansions of the rational
function $(z-w)^{-1-j}$ by
\begin{align*}
\big( \frac{1}{(z-w)^{1+j}} \big)_{\Expand{z}{w}} \quad = \quad
    \sum_{m=0}^\infty \binom{m}{j} z^{-1-m} w^{m-j}
    & \qquad \in \bC[[z^{-1},w]] \\
\big( \frac{1}{(z-w)^{1+j}} \big)_{\Expand{w}{z}} \quad = \quad
    - \sum_{m=-\infty}^{-1} \binom{m}{j} z^{-1-m} w^{m-j}
    & \qquad \in \bC[[z,w^{-1}]]
\textrm{ .}
\end{align*}
The formal delta function
$\delta(z-w) = \sum_{n \in \bZ} z^{-1-n} w^{n}
\in \bC[[z,z^{-1},w,w^{-1}]]$ and its derivatives are differences of
two expansions
\begin{align*}
\frac{1}{j!} \partial_w^j \delta(z-w) =
    \big( \frac{1}{(z-w)^{1+j}} \big)_{\Expand{z}{w}}
    - \big( \frac{1}{(z-w)^{1+j}} \big)_{\Expand{w}{z}}
\textrm{ .}
\end{align*}
The delta function has the following important property
\begin{align*}
\Res{z} h(z) \delta(z-w) = h(w)
\end{align*}
for all $h(z) \in U[[z,z^{-1}]]$. This is an analogue of a basic
result for analytic functions, where residues can be taken by
contour integration. If $h(z)$ is holomorphic then the difference
of contour integrals around origin of $h(z)/(z-w)$, for $|z|$ big
and $|z|$ small, is seen by contour deformation to correspond to the
residue at $z=w$, that is $h(w)$.

\bigskip

For the rest of the paper we will denote
\begin{align*}
f(z) = z + f_{-2} z^{-1} + f_{-3} z^{-2} + \cdots
    = z + \sum_{m \leq -2} f_m z^{1+m} \quad \in z \bC[[z^{-1}]]
\end{align*}
the formal 
distribution analogue of hydrodynamically normalized
conformal map, (\ref{eq: g coefs}).
Rather naturally we also use the formal distributions which are
expansions at infinity of quantities like
\begin{align*}
f'(z) \quad = \quad& 1 - f_{-2} z^{-2} - 2 f_{-3} z^{-3}
    - 3 f_{-4} z^{-4} - \cdots \\
f(z)^n \quad = \quad & z^n + n f_{-2} z^{n-2} + n f_{-3} z^{n-3}
    + ( n f_{-4} + \binom{n}{2} f_{-2}^2) z^{n-4} + \cdots \\
Sf(z) \quad = \quad & 
    - 6 f_{-2} z^{-4} - 24 f_{-3} z^{-5} - (60 f_{-4} + 12 f_{-2}^2)
    z^{-6} 
    + \cdots 
\textrm{ ,}
\end{align*}
all in the space $\bC((z^{-1}))$ of formal Laurent series at
$\infty$. Note that products of these series are well defined in
$\bC((z^{-1}))$.

\subsection{Homogeneity}
\label{sec: homogeneity}
Let us introduce a homogeneity degree that clarifies the algebraic
manipulations
and has a concrete geometric meaning. If one was to scale the SLE
hull by a factor $\lambda>0$, $\tilde{K}_t = \lambda K_t$, one
would end up with the uniformizing map
$\tilde{g}_t(z) = \lambda g_t(z/\lambda)$, i.e.
$\tilde{g}_m = \lambda^{-m} g_{m}$, driving processes
$\tilde{X}^I_t = \lambda X^I_t$, passive points
$\tilde{Y}^K_t = \lambda Y^K_t$ and growth speeds
$\ud \qvl \tilde{A}^I \qvr_t = \lambda^2 \, \ud \qvl A^I \qvr_t$.
We don't care so much of the
change in speeds, but we assign a degree $1$ to variables
$x_1, \ldots, x_N; y_1, \ldots, y_M$ and a
degree $-m$ to $f_{m}$. An element
$\phi \in \sF$ is called homogeneous of degree $\Delta$ if
$\phi(\lambda x_1, \ldots, \lambda y_M ; \lambda^2 f_{-2},
\lambda^3 f_{-3}, \ldots) =
\lambda^\Delta \, \phi(x_1,\ldots,y_M; f_{-2}, f_{-3}, \ldots)$.
Such scaling arguments are useful in figuring out how results look
in general. The simplest cases are e.g.
$\tilde{f}(z)^n = \lambda^n f(z/\lambda)^n$, which tells us that the
coefficient of $z^k$ in the expansion of $f(z)^n$ must be a polynomial
of degree $n-k$ in the $f_l$. Another example is derivatives, 
$\tilde{f}^{(m)}(z) = \lambda^{1-m} f^{(m)}(z/\lambda)$ so that the
coefficient of $z^k$ in the expansion of $f^{(m)}(z)$ must be of
degree $1-m-k$.

\bigskip

Rational functions of $f$ deserve a comment. Note that whenever
$F \in \bC((z^{-1}))$, we can ``compose'' $F(f(z)) \in \bC((z^{-1}))$
by using $f(z)^n \in z^n \bC[[z^{-1}]]$ --- only finitely many terms
contribute to a fixed power of $z$. Thus the notation of rational
functions of $f$ means that we first expand the rational function and
then the $f(z)^n$ terms, e.g.
\begin{align*}
\Big( \frac{1}{(f(z)-x)^{1+j}} \Big)_{\Expand{f(z)}{x}}
    = \sum_{m=0}^\infty \binom{m}{j} f(z)^{-1-m} x^{m-j}
\quad \in \bC[[z^{-1}]]((x))
\textrm{ .}
\end{align*}
The coefficient of $z^k$ is homogeneous of degree $-1-j-k$ since $x$
is of degree $1$.
We often need to replace $x$ in the above expression by $f(w)$, say.
But this still makes perfect sense in
$\big( \bC((w^{-1})) \big) [[z^{-1}]]$.

We will furthermore record for future application a ``change of
variables formula''
\begin{align}
\label{eq: change of variables}
\Res{z} f'(z) F(f(z)) = \Res{z} F(z)
\textrm{ .}
\end{align}
To prove the formula it is by linearity enough to prove it for
$F(z)=z^n$, that is $\Res{z} f'(z) f(z)^n = \delta_{n,-1}$.
For $n \neq -1$ we can use Leibniz's rule
$f'(z) f(z)^n = \frac{1}{n+1} \partial_z \big( f(z)^{1+n} \big)$
so the residue vanishes. For $n=-1$ on the other hand one has
$f'(z) f(z)^{-1} = \big( 1 + \Order (z^{-2}) \big) \big( z^{-1}
+ \Order(z^{-2}) \big)
= z^{-1} + \Order(z^{-2})$ so the residue is equal to $1$.

\subsection{The representation of $\vir$ on $\sF$}
\label{sec: the representation}

We are now ready to check that the formula (\ref{eq: Ln naive formula})
defines a representation of Virasoro algebra on $\sF$.
Working with formal series we ought to indicate carefully the
expansions we use and therefore the proper definition reads
\begin{align}
\nonumber
& \qquad \sL_n \quad = \\
\nonumber
& \Res{u} u^{1-n} \bigg\{
    \frac{c}{12} \; Sf(u) 
- \sum_{l \leq 2} \Res{z} f'(u)^2 z^{-2-l} 
    \big( \frac{1}{f(z) - f(u)} \big)_{\Expand{f(z)}{f(u)}} \;
    \frac{\partial}{\partial f_l} \\
\nonumber
& + \sum_{I} f'(u)^2 \Big(
    \delta_{x_I} \big( \frac{1}{(f(u)-x_I)^2} \big)_{\Expand{f(u)}{x_I}} 
    + \big( \frac{1}{f(u)-x_I} \big)_{\Expand{f(u)}{x_I}}
    \; \pder{x_I} \Big) \\
\label{eq: Ln formula}
& + \sum_{K} f'(u)^2 \Big(
    \delta_{y_K} \big( \frac{1}{(f(u)-y_K)^2} \big)_{\Expand{f(u)}{y_K}}
    + \big( \frac{1}{f(u)-y_K} \big)_{\Expand{f(u)}{y_K}}
    \; \pder{y_K} \Big) \bigg\}
\textrm{ .}
\end{align}
The numbers $c, \delta_{x_I}, \delta_{y_K} \in \bC$ are free parameters
so far.
But for the representation to be of relevance for SLE the parameters
will have to take specific values, see Proposition \ref{prop: A}.

We remark that all terms are either multiplication operators by
polynomials in $x_I$, $y_K$, $f_l$, or a derivative in one of
the variables composed with a multiplication by polynomial.
Therefore $\sL_n$ are clearly well defined on $\sF$.

A more detailed look at the polynomials reveals that each of these
is homogeneous and the degrees are such that $\sL_n$ lowers the
degree of a function by $n$.
Explicit expressions for $\sL_n$, $n \geq -2$, are listed in Appendix
\ref{app: explicit L}.

It is convenient to form a generating function of $L_n$:
the stress tensor, formally 
$T(\zeta) = \sum_{n \in \bZ} \zeta^{-2-n} L_n$. For our representation
defined by (\ref{eq: Ln formula}) we have
$\sT (\zeta) = \sum_n \zeta^{-2-n} \sL_n \in
(\End \sF) [[\zeta, \zeta^{-1}]]$ given explicitly by
\begin{align}
\nonumber
\sT (\zeta) \quad = & \quad \frac{c}{12} \zeta^{-4} Sf(\frac{1}{\zeta}) \\
\nonumber
& + \zeta^{-4} f'(\frac{1}{\zeta})^2 \; \bigg\{
    - \sum_{l\leq -2} \Res{w} w^{-2-l} \big( \frac{1}{f(w)-f(\frac{1}{\zeta})}
    \big)_{\Expand{f(w)}{f(\frac{1}{\zeta})}}
    \; \frac{\partial}{\partial f_l} \\
\nonumber
& + \sum_I \Big( \big(\frac{1}{f(\frac{1}{\zeta}) - x_I}
    \big)_{\Expand{f(\frac{1}{\zeta})}{x_I}}
    \; \frac{\partial}{\partial x_I}
    + \delta_{x_I} \big(\frac{1}{(f(\frac{1}{\zeta}) - x_I)^2}
    \big)_{\Expand{f(\frac{1}{\zeta})}{x_I}} \Big) \\
\label{eq: stress tensor}
& + \sum_K \Big( \big(\frac{1}{f(\frac{1}{\zeta}) - y_K}
    \big)_{\Expand{f(\frac{1}{\zeta})}{y_K}}
    \; \frac{\partial}{\partial y_K}
    + \delta_{y_K} \big(\frac{1}{(f(\frac{1}{\zeta}) - y_K)^2}
    \big)_{\Expand{f(\frac{1}{\zeta})}{y_K}} \Big)
\textrm{ .}
\end{align}
The $\sL_n$ are recovered as
$\sL_n = \Res{\zeta} \zeta^{1+n} \sT(\zeta)
= \Res{u} u^{-3-n} \sT(u^{-1})$.
\begin{remark}
As expected, $\sT(\zeta)$ is ``located'' in the physical space at
$1/\zeta$. This is because the $\sL_n$ morally act on the
contravariant module and produce descendants of the operator at
infinity, see Section \ref{sec: role of M}.
\end{remark}

To show that formula (\ref{eq: Ln formula}) defines a representation
of $\vir$ it is slightly more convenient to compute the following
commutator.
\begin{proposition}
\label{prop: Virasoro}
We have the following commutation relation
\begin{align*}
[\sL_n , \sT(\zeta)] = \frac{c}{12} (n^3-n) \zeta^{n-2}
    + 2 \sT(\zeta) (1+n) \zeta^{n} + \sT'(\zeta) \zeta^{1+n}
\textrm{ .}
\end{align*}
\end{proposition}
The computation is quite lengthy so we will give it in Appendix
\ref{app: Virasoro}.
Equivalent formulations of Proposition \ref{prop: Virasoro}
are given below, see e.g. \cite{Kac-vertex_algebras} Theorem 2.3.
\begin{corollary}
\label{cor: Virasoro}
The commutation relations of $\sL_n$ are
\begin{align*}
[\sL_n, \sL_m] = (n-m) \sL_{n+m} + \frac{c}{12} (n^3-n) \delta_{n+m,0}
\end{align*}
and thus they form a representation of $\vir$ on $\sF$.
Equivalently, we have the operator product expansion
\begin{align*}
[\sT(\zeta), \sT(\xi)] \quad = \quad & \frac{1}{3!} \partial_\xi^3
    \delta(\zeta-\xi) \; \Big( \frac{c}{2} \Big)
    + \partial_\xi \delta(\zeta-\xi) \; \Big( 2 \sT(\zeta) \Big) \\
& + \delta(\zeta-\xi) \; \Big( \sT'(\zeta) \Big)
\textrm{ .}
\end{align*}
\end{corollary}

\subsection{Local martingales}
\label{sec: local mgales}
Having defined a representation of $\vir$ in $\sF$, we now turn
to the topic of SLE local martingales.
As suggested in Section \ref{sec: SLE data} we pose the question for
which $\eta \in \sF$,
\begin{align}
\label{eq: eta over Z}
\frac{\eta(X^1_t, \ldots, X^N_t ; Y^1_t, \ldots, Y^M_t;
g_{-2}^{(t)}, g_{-3}^{(t)}, \ldots)}{Z(X^1_t, \ldots, X^N_t;
    Y^1_t, \ldots, Y^M_t)}
\end{align}
is a local martingale. The answer is given by Lemma
\ref{lem: drift operator}.
The notation will be simplified if we define for $I=1,\ldots,N$
the differential operators
\begin{align*}
\sA_I = \sD_I +
2 \sum_{m \leq -2} p_m(-x_I, f_{-2}, f_{-3}, \ldots) \pder{f_m}
\end{align*}
acting on $\sF$, where $\sD_I$ is as in (b) and
$p_m(f_{-1},f_{-2}, f_{-3}, \ldots)$
is the homogeneous polynomial\footnote{The $p_m$ were
also present in \cite{BB-SLE_martingales}. It is sometimes nice to
know that they can be obtained from the recursion $p_{-2} = 1$, and
$p_{-m} = - \sum_{k=1}^{m-2} f_{-k} \; p_{-m+k} $ for $m \leq -3$. Thus
e.g. $p_{-3} = -f_{-1}$, $p_{-4} = -f_{-2}+f_{-1}^2$ and so on.}
$\Res{v} v^{-2-m} (\frac{1}{f(v)+f_{-1}})_{\Expand{f(v)}{f_{-1}}}$
of degree $-2-m$.
\begin{lemma}
\label{lem: drift operator}
Suppose the SLE has driving processes and passive points
(\ref{eq: driving processes}), auxiliary function $Z$ satisfying (b) 
and $g_t$ defined by (\ref{eq: Loewner}) with coefficients
denoted as in (\ref{eq: g coefs}).
Then, for any $\eta \in \sF$, the It\^o drift of
(\ref{eq: eta over Z}) is given by
\begin{align*}
\frac{1}{Z(X^{1}_t, \ldots, Y^M_t)} \sum_{I=1}^N 
    \big( \sA_I \eta \big)
    (X^{1}_t, \ldots, 
    Y^M_t; g_{-2}^{(t)}, g_{-3}^{(t)} \ldots) \; \ud \qvl A^I \qvr_t 
\textrm{ .}
\end{align*}
\end{lemma}
\begin{Proof}
Observe that equation (\ref{eq: Loewner}) leads, by considering
$\oint_\infty \ud z \; z^{-2-m} \frac{\ud}{\ud t} g_t(z)$,
to the following drifts of the coefficients of $g_t$
\begin{align*}
\ud g_m^{(t)} \; = \; & 2 \sum_I \Res{z} z^{-2-m}
    \frac{1}{g_t(z)-X_t^{I}} \; \ud \qvl A^I \qvr_t \\
= \; & 2 \sum_I p_{m}(-X^{I}_t,g_{-2}^{(t)},g_{-3}^{(t)},\ldots)
    \; \ud \qvl A^I \qvr_t
\textrm{ .}
\end{align*}
The arguments of $\eta$ and $Z$ are governed by the above and the
diffusions (\ref{eq: driving processes}) so we can compute the drift of
$\eta / Z$ directly by It\^o's formula with the result
\begin{align*}
\; & \sum_I \Big( \ud \qvl A^I \qvr_t \kappa_I \frac{\pder{x_I} Z}{Z}
    + \sum_{J \neq I} \frac{2 \ud \qvl A^J \qvr_t}{x_I-x_J} \Big) \;
    \pder{x_I} \big( \frac{\eta}{Z} \big) \\
& + \sum_I \frac{\kappa_I \ud \qvl A^{I} \qvr_t}{2}
    \frac{\partial^2}{\partial x_I^2} \big( \frac{\eta}{Z} \big)
    + \sum_K \Big( \sum_J \frac{2 \ud \qvl A^J \qvr_t}{y_K-x_J} \Big)
    \; \pder{y_K} \big( \frac{\eta}{Z} \big) \\
& + \sum_{m \leq -2} \sum_I 2 \ud \qvl A^I \qvr_t \;
    p_m(-x_I, f_{-2}, \ldots) \;
    \pder{f_m} \big( \frac{\eta}{Z} \big) \\
= \;  & \sum_I \Big\{ \frac{\eta}{Z^2} \big( -\frac{\kappa_I}{2}
    \frac{\partial^2}{\partial x_I^2} Z - \sum_{J \neq I} \frac{2}{x_J-x_I}
    \; \pder{x_J} Z - \sum_K \frac{2}{y_K-x_I} \; \pder{y_K} Z \big) \\
& \qquad + \frac{1}{Z} \big(
    \frac{\kappa_I}{2} \frac{\partial^2}{\partial x_I^2} \eta
    + \sum_{J \neq I} \frac{2}{x_J-x_I} \pder{x_J} \eta
    + \sum_{K} \frac{2}{y_K-x_I} \pder{y_K} \eta  \\
& \qquad + 2 \sum_{m \leq -2} p_{m}(-x_I,\ldots) \; \pder{f_m} \eta \big)
    \Big\} \; \ud \qvl A^I \qvr_t
\textrm{ .}
\end{align*}
Now use the null field equation (b) to rewrite the
$\frac{\eta}{Z^2}$-term as
\begin{align*}
\frac{\eta}{Z} \big(
    \sum_{J \neq I} \frac{(\kappa_J-6)/\kappa_J}{(x_I-x_J)^2}
    - \sum_{K} \frac{2 h_{y_K}}{(y_K-x_I)^2} \big)
\textrm{ .}
\end{align*}
The assertion follows.
\end{Proof}
By Lemma \ref{lem: drift operator}, the operator $\sA_I$
corresponds to drift caused by growing the
$I^{\textrm{th}}$ curve. The crucial property of $\sA_I$, a
generalization of a result in \cite{BB-SLE_martingales},
is stated in the next Proposition and Corollary. The proof of the
Proposition is again left to Appendix \ref{app: T commutator A}.
Note that for these results we need to fix the values of the
parameters $\delta_{x_I}$, $\delta_{y_K}$ and $c$.
\begin{proposition} \label{prop: A}
If $\delta_{x_I} = h_{x_I} = \frac{6-\kappa_I}{2\kappa_I}$ for
all $I=1,\ldots,N$, $\delta_{y_K} = h_{y_K}$ for all $K=1,\ldots,M$
and $c = c(\kappa) = \frac{(6-\kappa)(3\kappa-8)}{2\kappa}$
we have
\begin{align*}
[\sT(\zeta), \sA_I] = -2 \zeta^{-4} f'(1/\zeta)^2 \big( \frac{1}
    {(f(1/\zeta)-x_I)^2} \big)_{\Expand{f(1/\zeta)}{x_I}} \; \sA_I
\textrm{ .}
\end{align*}
\end{proposition}
\begin{corollary} \label{cor: A}
If $\delta_{x_I}=h_{x_I}$,
$\delta_{y_K} = h_{y_K}$ and $c=c(\kappa)$
as in Proposition \ref{prop: A} we have
\begin{align*}
[\sL_n, \sA_I] = q_n (x_I; f_{-2}, f_{-3}, \ldots) \; \sA_I
\end{align*}
where $q_n$ is a homogeneous polynomial of degree $-n$, non-zero only
for $n \leq 0$. In particular, if $\eta \in \Kern \sA_I$, we have
$\sL_n \eta \in \Kern \sA_I$, too.
\end{corollary}
\begin{Proof}
Multiply the formula in Proposition \ref{prop: A} by $\zeta^{1+n}$ and
take the $\zeta$ residue to get\footnote{For concreteness, the
lowest $q_n$ are $q_0 = -2$, $q_{-1} = -4 x_I$,
$q_{-2} = -6 x_I^2 + 8 f_{-2}$.
As in \cite{BB-SLE_martingales}, it is possible to recover the higher
$q_{-m}$ from these recursively. We use
$L_{-n-1} = \frac{1}{n-1} [L_{-1}, L_{-n}]$ and Jacobi identity
to obtain for $n \geq 2$
\begin{align*}
[ \sL_{-n-1}, \sA_I ] \; = \; &
    \frac{1}{n-1} \big[ [\sL_{-1} , \sL_{-n}] , \sA_I \big] \\
= \; & \frac{1}{n-1} \big( - \big[ q_{-n}(x_I, f_{-2}, \ldots) \sA_I ,
    \sL_{-1} \big] + \big[ q_{-1}(x_I, f_{-2}, \ldots) \sA_I ,
    \sL_{-n} \big] \big) \\
= \; & \frac{1}{n-1} \big( [\sL_{-1}, q_{-n}(x_I, f_{-2},\ldots)]
    - [\sL_{-n}, q_{-1}(x_I) ] \big) \; \sA_I
\textrm{ .}
\end{align*} }
\begin{align*}
q_n(x_I;f_{-2}, f_{-3},\ldots) = -2 \; \Res{\zeta} \zeta^{-3+n}
    f'(1/\zeta)^2 \big( \frac{1}{(f(1/\zeta)-x_I)^2}
    \big)_{\expand{f(1/\zeta)}{large}}
\textrm{ .}
\end{align*}
The degree of homogeneity is easily found for example by comparing
\begin{align*}
-2 \zeta^{-4} \frac{f'(\frac{1}{\zeta})^2}{(f(\frac{1}{\zeta})-x)^{2}} = &
    \sum_m \zeta^m q_{-2-m}(x,f_{-2},\ldots) \qquad & \textrm{ and } \\
-2 \zeta^{-4} \frac{\til{f}'(\frac{1}{\zeta})^2}
    {(\til{f}(\frac{1}{\zeta})-\til{x})^{2}} = &
    \lambda^2 (\lambda \zeta)^{-4} f'(\frac{1}{\lambda \zeta})^2
    (f(\frac{1}{\lambda \zeta})-x)^{-2} \\
= & \sum_m \lambda^2 (\lambda \zeta)^m
    q_{-2-m}(x, f_{-2},\ldots)
\textrm{ .}
\end{align*}
The other claims are immediate consequences.
\end{Proof}

\begin{remark}
From now on we will always use the representation with the
values $c, \delta_{x_I}, \delta_{y_K}$ fixed in Proposition \ref{prop: A}
and Corollary \ref{cor: A} since it is the one that is useful for
building local martingales.
\end{remark}

\bigskip

In view of Lemma \ref{lem: drift operator} the subspace of $\sF$
of local martingales for the SLE is
$\{ \eta / Z \, : \, \eta \in \cap_I \Kern \sA_I \}$
and by Corollary \ref{cor: A}, $\cap_I \Kern \sA_I$ is a Virasoro module.
There is a submodule of great importance that can be
constructed from the partition function only, the one whose
motivation was discussed in Section \ref{sec: role of M}.
The partition function $Z \in \sF$ is a constant polynomial in the
variables $f_{-2}, f_{-3}, \ldots$ so the null field equations (b)
imply that $Z \in \cap_I \Kern \sA_I$. This is of course nothing
else but the trivial observation that the constant $Z / Z = 1$
is a local martingale. By Corollary \ref{cor: A} we can apply
the Virasoro generators to $Z$ to construct the space
\begin{align*}
\sM := \sU(\vir) Z \subset \cap_I \Kern \sA_I \subset \sF
\textrm{ .}
\end{align*}
Thus we have built a large amount of local martingales using only the
objects given by the definition of the SLE.

Let us state a couple of further easy consequences.
\begin{corollary}
Both $\sM$ and $\cap_I \Kern \sA_I$ are submodules of the
$\vir$-module $\sF$ and $\sM \subset \cap_I \Kern \sA_I$.
The auxiliary function $Z$ is annihilated
by $\sL_n$ for $n \geq 2$ and thus $\sM$ is spanned by
$\sL_{-m}^{n_{-m}} \cdots \sL_1^{n_1} Z$, where $m \geq -1$, $n_j \in \bN$
for all $j=-m,\ldots,1$. If we assume (c), then
$\sL_1 Z = 0$ and if we assume (d), then
$\sL_0 Z = (\Delta + \sum_I h_{x_I} + \sum_K h_{y_K}) Z$. In conclusion, assuming (c) and (d),
$\sM$ is a highest weight module for $\vir$ with highest weight vector
$Z$ and highest weight $\Delta + \sum_I h_{x_I} + \sum_K h_{y_K}$.
\end{corollary}
\begin{Proof}
That $\sF$ has submodule $\cap_I \Kern \sA_I$ was shown in Corollary
\ref{cor: A}. We observed that $Z \in \cap_I \Kern \sA_I$ and
defined $\sM$ as the minimal submodule containing $Z$.
The explicit expressions for $\sL_n$ given in
Appendix \ref{app: explicit L} show that $\sL_n$ $n \geq 2$
contain only terms $(\cdots) \pder{f_l}$ and thus they annihilate
functions that don't depend on $f_{-2}, f_{-3}, \ldots$, in particular
$Z$. The only term in $\sL_1$ that is not of this form is
$\sum_I \pder{x_I} + \sum_K \pder{y_K}$ and the only such term in
$\sL_0$ is
$\sum_I (\delta_{x_I} + x_I \pder{x_I}) +
\sum_K (\delta_{y_K} + y_K \pder{y}_K)$.
Thus the assumption (c) of translation invariance guarantees
$\sL_1 Z = 0$ and the assumption (d) of homogeneity gives
$\sL_0 Z = (\Delta + \sum_I h_{x_I} + \sum_K h_{y_K}) Z$.
\end{Proof}


\begin{remark}
\label{rem: acting directly}
It may seem slightly inconvenient that we have chosen a
representation which preserves the space of
``$Z$ times local martingales'' and not local martingales
themselves. If we have a local martingale $\vphi \in \sF$,
then $(\sL_n (Z \vphi)) / Z$ is another local martingale.
It would of course be possible to redefine
$\hat{\sL}_n \vphi = (\sL_n (Z \vphi)) / Z$. The $\hat{\sL}_n$
define a representation of $\vir$ and they now preserve the kernel
of the generator of our diffusion.
The local martingales corresponding to $\sM$ are those obtained
by repeated action of $\hat{\sL}_n$ on constant function $1$.
But the formula has become $Z$-dependent
\begin{align*}
\hat{\sL}_n \; = \; \sL_n + \sum_I \frac{\pder{x_I} Z}{Z}
    \Res{u} \frac{u^{1-n} f'(u)^2}{f(u) - x_I}
    + \sum_K \frac{\pder{y_K} Z}{Z}
    \Res{u} \frac{u^{1-n} f'(u)^2}{f(u) - y_K}
\end{align*}
expanded in $|f(u)|>|x_I|$ and $|f(u)|>|y_K|$.
\end{remark}

\subsection{First examples}
\label{sec: first examples}
\subsubsection{The chordal SLE}
The simplest SLE variant, chordal SLE, is a random curve from one boundary
point of a domain to another. It is customary to choose the domain to
be the half-plane $\bH$, starting point of the curve the origin $X_0=0$
and end point infinity. The number of curves is one, $N=1$, and there
are no other marked points $M=0$. The partition function is a constant
$Z(x)=1$. It is also customary to fix the time parametrization by
$\qvl A \qvr_t = t$.

The local martingales of chordal SLE were studied in
\cite{BB-SLE_martingales}. Due to constant $Z$, the operator $\sA$
is just the generator of the diffusion in variables
$X_t, g_{-2}^{(t)}, g_{-3}^{(t)}, \ldots$ and $X_t$ is merely
a Brownian motion with variance parameter $\kappa$. It is then
possible to consider $\sA$ as an operator on
the space of polynomials $\bC [x,f_{-2},f_{-3},\ldots]$.
It was shown that $\Kern \sA \subset \bC[x,f_{-2},f_{-3},\ldots]$
is a Virasoro module with central charge $c = c(\kappa)$ and constant
functions having $\sL_0$ eigenvalue
$h_{1,2}(\kappa) = \frac{6-\kappa}{2 \kappa}$.
Moreover, the fact that polynomials form a
vector space graded by integer degree
(defined as in Section \ref{sec: homogeneity}) such that the
subspaces are finite dimensional allowed a clever argument to
show that the graded dimension of
$\Kern \sA \subset \bC[x,f_{-2},\ldots]$
is precisely that of a generic irreducible highest weight module
degenerate at level two. Consequently for generic\footnote{The word
\emph{generic} here refers to Feigin-Fuchs Theorem about the submodule
structure of Verma modules for the Virasoro algebra
\cite{FF-Verma_modules_1983, FF-Verma_modules_1984,
FF-representations}. Thus generic
$\kappa$ means simply $\kappa \notin \bQ$: any highest weight module
of central charge $c(\kappa)$ is then either irreducible or contains
exactly one nontrivial submodule, which in turn is an irreducible
Verma module.
For $\kappa \in \bQ$ the situation may well be more complicated.
} $\kappa$,
the space of polynomial local martingales forms the irreducible highest
weight module of highest weight $h_{1,2}$.

Another viewpoint to the chordal SLE case is the verification in
\cite{BB-CFTs_of_SLEs, BB-conformal_transformations} that
the SLE state $G_{h_t} \omega_{1,2}$ is a local martingale,
where $h_t(z) = g_t(z)-X_t$ and $\omega_{1,2}$ is a highest
weight vector in the quotient of Verma module
$\Verma_{c(\kappa), h_{1,2}(\kappa)}$ by submodule generated by
the singular vector at level two.
Actually,
this completely solves the question of the structure of $\sM$ ---
the contravariant module of any highest weight representation is
a direct sum of irreducible highest weight representations, so $\sM$
is the irreducible highest weight representation with highest weight
$h_{1,2}$.
As a consequence for certain values of $\kappa$, the module $\sM$
generated by action of Virasoro generators on constant functions
is not the whole kernel, $\sM \subsetneq \Kern \sA$.
Easiest such degeneracies are $\kappa = 6$, in which case in fact
$\sM$ consists solely of constant functions
(the irreducible highest weight module with $c=0$, $h=0$ is one dimensional),
and $\kappa \in \{ 3, 10 \}$ in which cases
there are local martingales of homogeneity degree $3$ that can not
be obtained by the action of Virasoro algebra on constants functions.

\subsubsection{A coordinate change of chordal SLE}
By a M\"obius coordinate change of the ordinary chordal SLE
one defines the chordal SLE in $\bH$ from $X_0$ to $Y_0$,
see e.g. \cite{SW-coordinate_changes}.
The resulting process is an SLE${}_\kappa(\rho)$ with $\rho = \kappa-6$,
which is the SLE variant with one curve $N=1$, one passive point
$M=1$ (the marked point is $Y_0$) and partition function
$Z(x,y) = (x-y)^{(\kappa-6)/\kappa}$.
The conformal weights are equal,
$h_x = h_y = h_{1,2}(\kappa) = \frac{6-\kappa}{2 \kappa}$.
We remark that this variant can also be seen as a special case of
the a particular double SLE ``pure geometry'', see Section
\ref{sec: pure geometries} and \cite{BBK-multiple_SLEs}.

The structure of the Virasoro module $\sM$
as well as other properties of this case are studied in more detail
in the articles \cite{KK-reversibility_duality, KKM-in_preparation}
about chordal SLE reversibility.
Here we make some remarks that clarify the differences to the case
chordal SLE towards $\infty$ and in particular give some
justification to the choice of the straightforward approach taken.

The module $\sM$ is a highest weight module of highest weight
$\Delta + h_x + h_y = 0 = h_{1,1}(\kappa)$
and a direct computation gives $\sL_{-1} Z = 0$.
This is a manifestation of the M\"obius invariance of the process,
see Section \ref{sec: Mobius invariance}. For $\kappa \notin \bQ$ the
Verma module $\Verma_{c(\kappa),h_{1,1}(\kappa)}$
contains a single nontrivial submodule
generated by $L_{-1} \hwvV_{c(\kappa),h_{1,1}(\kappa)}$
and consequently $\sM$ must be the
irreducible highest weight module.
For the sake of illustration, below are the nonvanishing local
martingales up to level $4$:
\begin{align*}
\frac{\sL_{-2} Z}{Z} \; = \; &
    h (y - x)^2 - \frac{c}{2} f_{-2} \\
\frac{\sL_{-3} Z}{Z} \; = \; &
    2 h (y-x)^2 (x+y) - 2 c f_{-3} \\
\frac{\sL_{-4} Z}{Z} \; = \; &
    h (y-x)^2 (3 x^2 + 4 x y +3 y^2 - 6 f_{-2})
    - c (f_{-2}^2 + 5f_{-4}) \\
\frac{\sL_{-2} \sL_{-2} Z}{Z} \; = \; &
    \frac{c}{2} (f_{-2}^2 - 6f_{-4}) + \big( h (y-x)^2
    - \frac{c}{2} f_{-2} \big)^2 \\
& \qquad + 2 h (y-x)^2
    \big( -4 f_{-2} + x^2 + x y +y^2 \big)
\textrm{ ,}
\end{align*}
where $h=h_{1,2}(\kappa)=\frac{6-\kappa}{2 \kappa}$ and
$c=c(\kappa)=\frac{(6-\kappa)(3\kappa-8)}{2 \kappa}$.
\begin{remark}
A simple application of the listed local martingales would be the
determination of expected value of final half plane capacity of the hull.
Let $\tau$ denote the stopping
time $\inf \{t \geq 0 : X_t = Y_t\}$. Then if 
the capacity $g_{-2}^{(\tau)}$ of $K_\tau$
is integrable, i.e. $g_{-2}^{(\tau)} \in L^1(\PR)$, the local
martingale at level $2$
\begin{eqnarray*}
(Y_t-X_t)^2 - \frac{3\kappa - 8}{2} g_{-2}^{(t)}
\end{eqnarray*}
is a closable martingale up to the stopping time $\tau$ and
\begin{eqnarray*}
\EX [ g_{-2}^{(\tau)}] = \frac{2}{8-3 \kappa}
    (Y_0 - X_0)^2
\textrm{ .}
\end{eqnarray*}
The capacity $g_{-2}^{(\tau)}$ is an almost surely positive
finite quantity so for $\kappa \geq 8/3$ it is certainly
not in $L^1(\PR)$!
\end{remark}

Let us take a closer look at some of the most degenerate cases to
illustrate what may happen for rational values of $\kappa$.
For $\kappa=6$ we have $c(\kappa)=0$, $h(\kappa)=0$ and we have
a null vector $\sL_{-2} Z = 0$ at level two: the
representation $\sM$ is then indeed the irreducible (one dimensional!)
highest weight representation. The same central charge $c(\kappa)=0$
is obtained also with $\kappa=8/3$. For $\kappa=8/3$ the vector
$\sL_{-2} Z = \frac{5}{8} \, (y-x)^2 \, Z$ is directly checked to be
a nonzero singular vector in view of explicit expressions in Appendix
\ref{app: explicit L}, so $\sM$ is reducible. It takes a little bit
more of work to check that $\sL_{-2} \sL_{-2} Z$ and $\sL_{-4} Z$
become linearly dependent at $\kappa=10$ and thus there is a null
vector at level four, whereas at $\kappa=8/5$ at level four there
exists a singular vector $(y-x)^4 \, Z$.

The observation that $\sM$ can be reducible shows in particular that
one can't always construct the SLE${}_\kappa(\rho)$ state in
the form $M_t = G_{g_t} \Psi(X_t, Y_t) \omega_0$ taking values in
a Virasoro highest weight module --- recall that the contravariant
module of a highest weight module is a direct sum of irreducible
representations so $\sM$ couldn't be a submodule of the contravariant
module. This shows an advantage of proceeding directly in the manner
of the whole Section \ref{sec: Virasoro structure}: we found
the representation $\sM$ without addressing the question of the
space in which to construct $M_t$!

Yet another thing that is well illustrated by this variant is the
fact that $\sM$ doesn't contain all local martingales even for
$\kappa$ generic.
The function $\zeta(x,y) = (y-x)^{2/\kappa}$ is also annihilated
by $\sA$.
Actually, $\zeta$ arises as another ``pure geometry'' of double SLE,
see Section \ref{sec: pure geometries}.
So for example the following are local
martingales for the SLE${}_\kappa (\kappa-6)$
\begin{align*}
& & \frac{\zeta}{Z} \; = \; (y-x)^{\frac{8-\kappa}{\kappa}} \\
& & \frac{\sL_{-1} \zeta}{Z} \; = \; \frac{8-\kappa}{\kappa} (y+x)
    \; (y-x)^{\frac{8-\kappa}{\kappa}} \\
& & \frac{\sL_{-2} \zeta}{Z} \; = \Big( 
    \frac{(3 \kappa^2 - 10 \kappa - 80) f_{-2}
    + (44-6\kappa)(x^2 + y^2) + 8 x y}{4 \kappa} \Big)
    \; (y-x)^{\frac{8-\kappa}{\kappa}}
\textrm{ .}
\end{align*}
These local martingales are not polynomial in $x_I$ and $y_K$.
Also, we know that $\sU(\vir) \zeta$ is a highest
weight module of highest weight $h_{1,3}$, so we get a
lot of local martingales not contained in $\sM$.
This shows another difference to the case of ordinary chordal SLE
towards infinity.



\subsubsection{SLE${}_\kappa (\rho)$}
The SLE variant SLE${}_\kappa (\rho_1,\rho_2,\ldots,\rho_M)$
has one curve $N=1$ and several marked points $M \geq 1$.
Its partition function is
\begin{align}
\label{eq: Z for kappa rho}
Z(x; y_1, \ldots, y_M) = \Big( \prod_{K=1}^M
    (y_K-x)^{\rho_K/\kappa} \Big) \Big( \prod_{1 \leq J < K \leq M}
    (y_J - y_K)^{\rho_J \rho_K / 2 \kappa} \Big)
\end{align}
as discussed in \cite{Kytola-SLE_kappa_rho}. The conformal
weights are
$\delta_{y_K} = \frac{1}{4 \kappa} \rho_K (\rho_K + 4 - \kappa)$.
The driving process satisfies 
$\ud X_t = \sqrt{\kappa} \; \ud A_t + 
\sum_K \frac{\rho_K}{X_t-Y^K_t} \; \ud \qvl A \qvr_t$,
which was taken as the definition when the variant was
introduced \cite{LSW-conformal_restriction, Dubedat-duality}.
It would also be natural to generalize
the definition to include possibility of bulk marked points,
see \cite{SW-coordinate_changes}.

Just like for SLE${}_\kappa(\kappa-6)$ or double SLE, the module
$\sM$ consists of elements that are polynomial times $Z$.
Thus the local martingales obtained this way are polynomial.
This becomes obvious in the light of Remark
\ref{rem: acting directly} and the partition function
(\ref{eq: Z for kappa rho}) since
\begin{align*}
& \Res{u} \Big( \frac{\pder{x} Z}{Z} \frac{u^{1-n} f'(u)^2}{f(u) - x}
    + \sum_K \frac{\pder{y_K} Z}{Z} \frac{u^{1-n} f'(u)^2}{f(u) - y_K} \Big) \\
= \; & \Res{u} \Big(
    \sum_K \frac{\frac{\rho_K}{\kappa} u^{1-n} f'(u)^2}{(f(u)-x)(f(u)-y_K)}
    + \sum_{K < K'} \frac{\frac{\rho_K \rho_{K'}}{2 \kappa} u^{1-n} f'(u)^2}
    {(f(u)-y_K)(f(u)-y_{K'})} \Big)
\end{align*}
so the additional multiplication operator in $\hat{\sL}_n$ is
polynomial.
Of course $\sM$ may not be the whole of $\Kern \sA$ and there may
be other local martingales that are non-polynomial.

As has been remarked by several authors
\cite{Cardy-SLE_kappa_rho, Kytola-SLE_kappa_rho, MRR-SLE_and_Coulomb_gas}, 
SLE${}_\kappa (\rho_1,\rho_2,\ldots,\rho_M)$ seems to be
best studied using the Coulomb gas formalism.
Indeed, after having discussed the Coulomb gas briefly,
we will show how to construct the SLE state $M_t$
conveniently in charged Fock spaces.

\begin{remark}
Articles \cite{Dubedat-duality, Dubedat-commutation} suggest that
a famous conjecture of SLE duality be formulated in a global fashion
using the SLE${}_\kappa (\underline{\rho})$ processes.
The structure of $\sM$ in these cases gives strong support
for the proposed approach, see \cite{KK-reversibility_duality}.
\end{remark}

\subsubsection{Multiple SLEs}
The multiple SLEs in the sense of \cite{BBK-multiple_SLEs}
correspond to boundary conditions where there are several
interfaces starting from the boundary of the domain, but
no other marked points. Thus the parameters are $N \geq 2$,
$M=0$ and $\kappa_I$ takes the same value
$0 < \kappa < 8$ for all $I=1,\ldots,N$.

The multiple SLEs in this sense were also studied in
\cite{Dubedat-Euler_integrals} and can be seen as a special
case of the general definition of multiple SLEs through commutation
requirement of \cite{Dubedat-commutation}.
A very instructive study of multiple SLEs with emphasis on
absolute continuity of probability measures can be found in
\cite{Graham-multiple_SLEs}. There, different curves are allowed
to have different $\kappa_I \in \{ \kappa, 16/\kappa \}$.

In Section \ref{sec: Coulomb gas} we construct the state $M_t$
of multiple SLEs in a charged Fock space using the Coulomb
gas formalism.
It has been conjectured in \cite{BBK-multiple_SLEs} that the
asymptotics of $Z$ as certain points come close to each other
are related to topological configuration of the multiple SLE
curves.
Coulomb gas or Feigin-Fuchs integrals appear to make the
choice of conformal blocks transparent for the ``pure
geometries''.


\section{Coulomb gas constructions of SLE states}
\label{sec: Coulomb gas}

\subsection{Preliminaries: charged Fock spaces as Virasoro modules}
The following method, known as Coulomb gas, has been described
for example in \cite{TK-Fock_space_representations,
Felder-BRST_approach, FFK-braid_matrices}
and Chapter 9 of \cite{DMS-CFT}.

The method is to study certain modules for the Heisenberg algebra
generated by $a_n$, $n \in \bZ$, with commutation relations
$[a_n, a_m] = 2 \delta_{n+m,0}$. The ``charged Fock space''
$F_\alpha$ is
\begin{align*}
F_\alpha = \bigoplus_{k=0}^\infty
    \bigoplus_{1\leq n_1 \leq \cdots \leq n_k}
    \bC a_{-n_1} \cdots a_{-n_k} v_\alpha
\end{align*}
where $v_\alpha$ is a vacuum:
$a_0 v_\alpha = 2 \alpha v_\alpha$ and $a_n v_\alpha = 0$ for $n>0$.
One defines a representation of Virasoro algebra on $F_\alpha$ by
\begin{align*}
L_n  = \frac{1}{4} \sum_{j} \no a_{n-j} a_j \no - \alpha_0 (n+1) a_n
\end{align*}
where $\no a_n a_m \no$ is $a_n a_m$ if $n \leq m$ and $a_m a_n$ otherwise.
Note that acting on basis vectors $a_{-n_1} \cdots a_{-n_k} v_\alpha$
the sum over $j$ has only finitely many non-zero terms because
$a_j a_{-n_1} \cdots a_{-n_k} v_\alpha = 0$ for $j > \sum_i n_i$.
The parameter $\alpha_0$ determines the central charge, $c=1-24\alpha_0^2$.

The $L_0$ eigenvalue of $v_\alpha$ is
$h(\alpha) = \alpha^2 - 2 \alpha_0 \alpha$ and the
charged Fock space is a direct sum of finite dimensional $L_0$
eigenspaces, $F_\alpha = \oplus_{m=0}^\infty ( F_\alpha )_m$, where
$( F_\alpha )_m$ corresponds to eigenvalue $h(\alpha)+m$.
The eigenspace $(F_\alpha)_m$ has a basis consisting of
$a_{-n_1} \cdots a_{-n_k} v_\alpha$, with $1 \leq n_1 \leq \cdots \leq n_k$
and $n_1+\cdots+n_k=m$.

The contravariant module is defined
as a direct sum of duals of the finite dimensional eigenspaces
\begin{align*}
F^*_\alpha = \bigoplus_{m=0}^\infty (F_\alpha)_m^*
 =  \bigoplus_{k=0}^\infty \bigoplus_{1 \leq n_1 \leq \cdots \leq n_k}
    \bC a_{n_1}\tp \cdots a_{n_k}\tp v_\alpha^*
\textrm{ ,}
\end{align*}
where $v_\alpha^* \in ( F_\alpha )^*_0$ is such that
$\bra v_\alpha^*, v_\alpha \ket = 1$. It becomes a Virasoro module in
the usual way: $L_{-n}\tp$ have the same commutation relations as the
generators of $\vir$. We have a
bilinear pairing $\bra \cdot , \cdot \ket : F_\alpha^* \times F_\alpha
\rightarrow \bC$. It is often necessary to allow infinite linear
combinations of the basis vectors, so denote
$\hat{F}_\alpha = \prod_{m=0}^\infty ( F_\alpha )_m$
and $\hat{F}^*_\alpha = \prod_{m=0}^\infty ( F_\alpha )^*_m$. The
bilinear pairing extends naturally to $F_\alpha^* \times \hat{F}_\alpha$
and $\hat{F}^*_\alpha \times F_\alpha$.

Let us still introduce a convenient notation for the charges $\alpha$.
First of all $\alpha_\pm = \pm (\sqrt{\kappa}/2)^{\pm 1}$ and
$\alpha_0 = \half (\alpha_+ + \alpha_-)$ relate the SLE parameter
$\kappa$ to the Coulomb gas formalism. Then let
$\alpha_{n,m} = \frac{1-n}{2} \alpha_+ + \frac{1-m}{2} \alpha_-$.
Finally, we use the shorthand notation $F_{n,m} = F_{\alpha_{n,m}}$,
$h_{n,m}=h(\alpha_{n,m})$ etc.

\subsection{Preliminaries: vertex operators and screening charges}
The Coulomb gas formalism constructs intertwining operators
between the charged Fock spaces morally as the normal ordered
exponentials $\no e^{\alpha \vphi(z)} \no$ of the free massless boson
field $\vphi$. More precisely,
$V_{\alpha}(z) : F_\beta \rightarrow \hat{F}_{\beta+\alpha}$ is
defined by
\begin{align*}
V_\alpha (z) \; = \; & z^{2 \alpha \beta} \;
    U^-_{\alpha}(z) \; U^+_{\alpha}(z) \; T_\alpha \; \textrm{ ,}
    & \quad \textrm{ where} \\
[a_n , T_\alpha] \; = \; & 2 \alpha \; \delta_{n,0}
\qquad , \qquad T_\alpha v_\beta \; = \; v_{\alpha+\beta}
& \quad \textrm{ and} \\
U^\pm_\alpha (z) \; = \; & \exp \big( \mp \sum_{n=1}^\infty \frac{1}{n}
    \alpha z^{\mp n} a_{\pm n} \big)
\end{align*}
for $z$ in the universal covering manifold of $\bC \setminus \{ 0 \}$.
Again the definition makes sense because only finitely many nonzero
terms are created by $U^+_\alpha (z)$ acting on any
$u \in F_{\beta+\alpha}$.
The vertex operators are intertwining operators of conformal weight
$h(\alpha)$
\begin{align*}
[L_n , V_\alpha (z)]
    = (z^{1+n} \pder{z} + (1+n) h(\alpha) z^n) \; V_\alpha (z)
\textrm{ .}
\end{align*}

There is a way to make sense of compositions of vertex operators
$V_{\alpha_1}(z_1) \cdots V_{\alpha_n}(z_n)$ in the region
$|z_1| > \cdots > |z_n|$, see e.g. \cite{TK-Fock_space_representations}.
The formula thus obtained can be analytically continued to the universal
covering manifold of $\{ (z_1, \ldots, z_n) \in \bC^n :
z_i \neq z_j \; \forall i,j \textrm{ and } z_i \neq 0 \; \forall i \}$,
\begin{align*}
V_{\alpha_1,\ldots,\alpha_n} (z_1, \ldots, z_n)
    \; = \; &  \; h_{\beta; \underline{\alpha}}(z_1, \ldots, z_n)
    U^-_{\underline{\alpha}}(z_1, \ldots, z_n) \;
    U^+_{\underline{\alpha}}(z_1, \ldots, z_n) \;
    T_{\sum_i \alpha_i} \\
h_{\beta; \alpha_1,\ldots,\alpha_n}(z_1, \ldots, z_n)
    \; = \; & \prod_{i=1}^n z_i^{2 \alpha_i \beta}
    \prod_{1 \leq i < j \leq n} (z_i - z_j)^{2 \alpha_i \alpha_j} \\
U^\pm_{\alpha_1,\ldots,\alpha_n} (z_1, \ldots, z_n)
    \; = \; & \exp \big( \mp \sum_{n=1}^\infty \frac{1}{n}
    (\sum_{i} \alpha_i z_i^{\mp n}) a_{\pm n} \big)
\end{align*}
and we take this $V_{\underline{\alpha}}(z_1, \ldots, z_n) :
F_\beta \rightarrow \hat{F}_{\beta + \alpha_1 + \cdots + \alpha_n}$
as the definition of composition of vertex operators.
We have the intertwining relation
\begin{align*}
[L_n , V_{\underline{\alpha}} (z_1, \ldots, z_n)]
    = \sum_i \big( z_i^{1+n} \pder{z_i} + (1+n) h(\alpha_i) z_i^n \big)
    \; V_{\underline{\alpha}} (z_1, \ldots, z_n)
\textrm{ .}
\end{align*}

To construct further intertwining operators, one makes the following
observation. There are two values of $\alpha$ for which $h(\alpha) = 1$,
namely $\alpha = \alpha_\pm$. For these values the commutators
$[L_n , V_{\alpha_\pm} (z)]$ are total derivatives
$\frac{\ud}{\ud z} ( z^{1+n} V_{\alpha_\pm} (z))$.
Integrating the composition of vertex operators
\begin{align*}
V_{\alpha_1, \ldots, \alpha_n; \alpha_-, \ldots, \alpha_-;
\alpha_+, \ldots, \alpha_+}
(z_1, \ldots, z_n; w^-_1, \ldots, w^-_{s^-} ; w^+_1, \ldots, w^+_{s^+})
\end{align*}
in variables
$w^-_1, \ldots, w^-_{s^-} ; w^+_1, \ldots, w^+_{s^+}$
over contours $\Gamma$ such that
\begin{align*}
h_{\alpha_1, \ldots, \alpha_n; \alpha_-, \ldots, \alpha_+}
(z_1, \ldots, z_n; w^-_1, \ldots, w^+_{s^+})
\end{align*}
takes the same value at the endpoints, one defines
\begin{align*}
& V^{\Gamma; s^-, s^+}_{\alpha_1, \ldots, \alpha_n} (z_1, \ldots, z_n) \\
= \; & \int_\Gamma V_{\alpha_1, \ldots, \alpha_n; \alpha_-, \ldots, \alpha_+}
    (z_1, \ldots, z_n; w^-_1, \ldots, w^+_{s^+}) \;
    \ud w^-_1 \cdots \ud w^+_{s^+}
\textrm{ .}
\end{align*}
Now
$V^{\Gamma; s^-, s^+}_{\alpha_1, \ldots, \alpha_n} (z_1, \ldots, z_n) :
F_\beta \rightarrow
\hat{F}_{\beta + \sum_i \alpha_i + s^- \alpha_- + s^+ \alpha_+}$
is again an intertwining operator
\begin{align*}
& [L_n , V^{\Gamma; s^-, s^+}_{\underline{\alpha}} (z_1, \ldots, z_n)] \\
\; = \; & \sum_i \big( z_i^{1+n} \pder{z_i} + (1+n) h(\alpha_i) z_i^n \big)
    \; V^{\Gamma; s^-, s^+}_{\underline{\alpha}} (z_1, \ldots, z_n)
\end{align*}
because the total derivatives vanish after integration. The contours
$\Gamma$ should of course be chosen in such a way that
$V^{\Gamma; s^-, s^+}_{\underline{\alpha}}$ is not zero.
The additional charges $\alpha_\pm$ whose position was integrated
over are called screening charges and the resulting operator
$V^{\Gamma; s^-,s^+}_{\underline{\alpha}}$
is called a screened vertex operator.

\subsection{Application to SLE${}_\kappa (\rho)$}
The Coulomb gas method allows to build the state of
SLE${}_\kappa(\underline{\rho})$ explicitly. An easy choice of
the values of $\alpha$ was suggested in
\cite{Kytola-SLE_kappa_rho}, namely at $x$ one uses
$\alpha = \alpha_{1,2} = \frac{1}{\sqrt{\kappa}}$
and at $y_K$ the choice is
$\alpha_K = \frac{\rho_K}{2 \sqrt{\kappa}}$.
Then we have for $\alpha_\infty = \alpha_{1,2} + \sum_K \alpha_K$
\begin{align*}
& \bra v^*_{\alpha_\infty} ,
    V_{\alpha; \alpha_1, \ldots, \alpha_M} (x; y_1, \ldots, y_M) v_0 \ket \\
= \; & h_{0; \alpha, \alpha_1, \ldots, \alpha_M} (x, y_1, \ldots, y_M) \\
= \; & \Big( \prod_{K=1}^M
    (x-y_K)^{\rho_K/\kappa} \Big) \Big( \prod_{1 \leq J < K \leq M}
    (y_K - y_J)^{\rho_J \rho_K / 2 \kappa} \Big) \\
= \; & \const \times Z(x; y_1, \ldots, y_K)
\textrm{ .}
\end{align*}

It is well known that $h_{\alpha; \alpha_1, \ldots, \alpha_M}$
satisfies the following null field equation, but we give the
proof here as this is the key property for application to
SLE and is a natural step towards Lemma
\ref{lem: null field multiple SLE}.
\begin{lemma}
\label{lem: null field CG}
For $\alpha = \alpha_{1,2} = 1/\sqrt{\kappa}$ we have the null field
equation
\begin{align*}
\Big( \frac{\kappa}{2} \; \frac{\partial^2}{\partial x^2} +
    \sum_{K=1}^M \big( \frac{2}{y_K - x} \pder{y_K}
    - \frac{2 \delta_K}{(y_K-x_I)^2} \big) \Big)
    h_{0; \alpha, \alpha_1, \ldots, \alpha_M} (x; y_1, \ldots, y_M) = 0
\textrm{ ,}
\end{align*}
where $\delta_K = h(\alpha_K) = \alpha_K^2 - (\frac{\sqrt{\kappa}}{2}-\frac{2}{\sqrt{\kappa}}) \alpha_K$.
\end{lemma}
\begin{Proof}
We need to compute the following terms
\begin{align*}
\ppder{x} h \; = \; & \pder{x} \big( h \times
    (\sum_K \frac{2 \alpha_K / \sqrt{\kappa}}{x-y_K}) \big) \\
= \; & h \times \Big(
    (\sum_K \frac{2 \alpha_K / \sqrt{\kappa}}{x-y_K})^2
    - \sum_K \frac{2 \alpha_K / \sqrt{\kappa}}{(x-y_K)^2} \Big) \\
= \; & h \times \Big(
    \sum_{\substack{K, K' \\ K \neq K'}}
    \frac{4 \alpha_K \alpha_{K'} / \kappa}{(x-y_K) (x-y_{K'})}
    + \sum_K \frac{4 \alpha_K^2/\kappa - 2 \alpha_K / \sqrt{\kappa}}
    {(x-y_K)^2} \Big)
\end{align*}
and
\begin{align*}
& \sum \frac{2}{y_K - x} \pder{y_K} h \\ \; = \; &
    h \times \Big( \sum_K \frac{2}{y_K-x}
    \big( \frac{2 \alpha_K / \sqrt{\kappa}}{y_K-x}
    + \sum_{K' \neq K} \frac{2 \alpha_K \alpha_{K'}}{y_K - y_{K'}}
    \big) \Big) \\
= \; & h \times \Big( \sum_K \frac{4 \alpha_K / \sqrt{\kappa}}{(x- y_K)^2}
    + \sum_{\substack{K, K' \\ K < K'}}
    \frac{4 \alpha_K \alpha_{K'}}{(y_K - y_{K'})} \big(
    \frac{1}{y_K-x} - \frac{1}{y_{K'}-x} \big) \Big) \\
= \; & h \times \Big( \sum_K \frac{4 \alpha_K / \sqrt{\kappa}}{(x- y_K)^2}
    + \sum_{\substack{K, K' \\ K \neq K'}}
    \frac{-2 \alpha_K \alpha_{K'}}{(x-y_K) (x-y_{K'})} \Big) 
\textrm{ .}
\end{align*}
Now it's easy to see that all $\sum_{K \neq K'}$ terms cancel
in the following
\begin{align*}
\big( \frac{\kappa}{2} \; \frac{\partial^2}{\partial x^2} +
    \sum_{K=1}^M \frac{2}{y_K - x} \pder{y_K} \big) \; h 
\; = \; h \times \Big( \sum_K
    \frac{2 \alpha_K^2 - \sqrt{\kappa} \alpha_K
      + 4 \alpha_K/\sqrt{\kappa}}{(x-y_K)^2} \Big)
\end{align*}
and one only needs to observe that
$\alpha_K^2 - \frac{\sqrt{\kappa}}{2} \alpha_K
+ \frac{2}{\sqrt{\kappa}} \alpha_K = \alpha_K (\alpha_K-2\alpha_0) =
h(\alpha_K)$
to reach the conclusion.
\end{Proof}

We're in fact ready to give a straightforward computation that the
SLE${}_\kappa (\underline{\rho})$ state
\begin{align*}
M_t := \frac{1}{Z (X_t; Y^1_t, \ldots, Y^M_t)} G_{g_t}
    V_{\alpha; \alpha_1, \ldots, \alpha_M} (X_t; Y^1_t, \ldots, Y^M_t) v_{0}
\end{align*}
is a $\hat{F}_\beta$ valued local martingale, where
$\beta = {\alpha + \sum_K \alpha_K}$.
This means that as we express it in the basis
$a_{-n_1} \cdots a_{-n_k} v_{\beta}$, the
coefficients of the basis vectors are local martingales.

We should check that $\sA$ annihilates $Z \times M_t$.
Recalling that
$Z = \const \times h_{0; \alpha, \alpha_1, \ldots, \alpha_M}$
and $U^+(\cdots) v_\beta = v_\beta$
the local martingale property of $M_t$ is the vanishing of
\begin{align*}
\sA \; \Big( h \times G_f U^-_{\alpha; \alpha_1, \ldots, \alpha_M}
    (x; y_1, \ldots, y_M) v_\beta \Big)
\textrm{ .}
\end{align*}
But $\sA = \sD + 2 \sum_{l \leq -2} p_l (-x, f_{-2}, f_{-3}, \cdots)
\pder{f_l}$
and $\sD$ annihilates $h$ by Lemma \ref{lem: null field CG}. Thus
we are left to check that
\begin{align}
\nonumber
& \Big\{ h \times \frac{\kappa}{2} \ppder{x}
    + \big( \pder{x} h \big) \times \kappa \pder{x}
    + h \times \sum_K \frac{2}{y_K-x} \pder{y_K} \\
\nonumber
& \qquad + h \times 2 \sum_{l \leq -2} p_l (-x; f_{-2}, \ldots) \pder{f_l}
    \Big\} \; G_f \; U^- (x; y_1, \ldots, y_M) \; v_\beta \\
\nonumber
= \; & h \times \Big\{ \frac{\kappa}{2} \ppder{x}
    + \kappa \sum_K \frac{2 \alpha \alpha_K}{x-y_K} \pder{x}
    + \sum_K \frac{2}{y_K-x} \pder{y_K} \\
\label{eq: what should vanish}
& \qquad + 2 \sum_{l \leq -2} p_l (-x; f_{-2}, \ldots) \pder{f_l}
    \Big\} \; G_f \; U^- (x; y_1, \ldots, y_M) \; v_\beta
\end{align}
vanishes. We will split the verification of this to pieces.

Let us start with the computation of
$2 \sum_l p_l \pder{f_l} G_f U^- v_\beta$.
First step is to use definition of $p_l$ and defining property
(\ref{eq: defining G}) of $G_f$ to rewrite
\begin{align*}
& \sum_{l \leq -2} p_l(-x, f_{-2}, \ldots) \pder{f_l} G_f \\
\; = \; & - \sum_{\substack{l \leq -2 \\ k \leq l}} \Res{v} \Res{w}
    \big( \frac{v^{-2-l}}{f(v)-x} \big)_{\Expand{f(v)}{x}}
    \frac{w^{1+l} f'(w)}{f(w)^{2+k}} G_f L_k \\
\; = \; & - \sum_{\substack{k \leq -2}} \Res{v} \Res{w}
    \big( \frac{1}{f(v)-x} \big)_{\Expand{f(v)}{x}}
    \big( \frac{1}{w-v} \big)_{\Expand{w}{v}}
    \frac{f'(w)}{f(w)^{2+k}} G_f L_k
\textrm{.}
\end{align*}
Then we can change the expansion using
$\big( \frac{1}{w-v} \big)_{\Expand{w}{v}} =
\big( \frac{1}{w-v} \big)_{\Expand{v}{w}} + \delta(w-v)$
and observe that expanded in $|v|>|w|$ the $v$
residue vanishes so the above is equal to
\begin{align*}
\; & - \sum_{\substack{k \leq -2}} \Big( 0 + \Res{w}
    \big( \frac{1}{f(w)-x} \big)_{\Expand{f(w)}{x}}
    \frac{f'(w)}{f(w)^{2+k}} \Big) G_f L_k
\textrm{ .}
\end{align*}
The change of variables formula (\ref{eq: change of variables}) yields
\begin{align*}
\sum_{l \leq -2} p_l(\cdots) \pder{f_l} G_f
\; = \; & - \sum_{\substack{k \leq -2}} \Res{z}
    \big( \frac{1}{z-x} \big)_{\Expand{z}{x}}
    z^{-2-k} G_f L_k \\
\; = \; & - G_f \sum_{\substack{k \leq -2}} x^{-2-k} L_k
\textrm{ .}
\end{align*}
Having simplified a little we will commute the $L_k$ to the
right of $U^-$. For $k < 0$ and $n > 0$ we have
$[L_k , a_{-n}] = n a_{k-n}$ and since $a_{k-n}$ commutes with
$a_{-n'}$, $n' > 0$ this leads to
\begin{align*}
[L_k , U^- (x; y_1, \ldots, y_M)] =
    \sum_{n=1}^\infty (\alpha x^n + \sum_K \alpha_K y_K^n) a_{k-n}
    \; U^- (x; y_1, \ldots, y_M)
\textrm{ .}
\end{align*}
Consequently, to commute $L_k$ to right we generate in
(\ref{eq: what should vanish}) the terms
\begin{align}
\nonumber
& -2 h \times G_f \; \sum_{k \leq -2} x^{-k-2}
    [L_k , U^- (x; y_1, \ldots, y_M)] v_\beta \\
\nonumber
= \; & -2 h \times G_f \; \sum_{k \leq -2} \sum_{n=1}^\infty x^{-k-2}
    (\alpha x^n + \sum_K \alpha_K y_K^n) \; a_{k-n} \;
    U^- (x; y_1, \ldots, y_M) v_\beta \\
\label{eq: term 1}
= \; & h \times G_f \; \sum_{m=3}^\infty \Big( -2 \alpha (m-2) x^{m-2} 
    -2 \sum_K \alpha_K \big( \sum_{j=0}^{m-3} x^j y_K^{m-2-j} \big)
    \Big) \; a_{-m} U^- \; v_\beta
\textrm{ .}
\end{align}
After commutation the $L_k$ act on $v_\beta$,
\begin{align*}
\nonumber
& -2 \sum_{k \leq -2} x^{-k-2} L_k v_\beta
= -2 \sum_{m = 2}^\infty x^{m-2} L_{-m} v_\beta \\
\nonumber
= \; & -2 \sum_{m=2}^\infty x^{m-2} 
    \big( \frac{1}{4} \sum_{i \in \bZ} \no a_{-m-i} a_i \no
    - \alpha_0 (1-m) a_{-m} \big) v_\beta
\textrm{ ,}
\end{align*}
which gives the contribution
\begin{align}
\nonumber
= \; & h \times G_f \; \sum_{m=2}^\infty x^{m-2} 
    \Big( - \frac{1}{2} \sum_{i = -m+1}^{-1} a_{-m-i} a_i \\
\label{eq: term 2}
& \qquad + \big( -2 \beta - (m-1) (\frac{\sqrt{\kappa}}{2}
    - \frac{2}{\sqrt{\kappa}}) \big) a_{-m} \Big)
    \; U^-(x; y_1, \ldots, y_M) \; v_\beta
\textrm{ .}
\end{align}

So far we've computed only the $f_l$ derivatives in
(\ref{eq: what should vanish})
but the rest will be simpler, because the operators involved
commute with each other. Apart from the last term that was already
treated, (\ref{eq: what should vanish}) is
\begin{align*}
& h \times \Big\{ \frac{\kappa}{2} \ppder{x}
    + \kappa \sum_{K} \frac{2 \alpha \alpha_K}{x-y_K} \pder{x}
    + \sum_K \frac{2}{y_K-x} \pder{y_K} \Big\}
    \; G_f \; U^-(x; y_1, \ldots y_M) \; v_\beta \\
= \; & h \times G_f \; \Big\{ \frac{\kappa}{2} \big(
    \sum_{n,n'=1}^\infty \alpha^2 x^{n+n'-2} a_{-n} a_{-n'}
    + \sum_{n=1}^\infty (n-1) \alpha x^{n-2} a_{-n} \big) \\
& \qquad + \kappa \sum_K \frac{2 \alpha \alpha_K}{x-y_K}
        \sum_{n=1}^\infty \alpha x^{n-1} a_{-n}
    + \sum_K \frac{2}{y_K-x} \sum_{n=1}^\infty \alpha_K y_K^{n-1} a_{-n}
    \Big\} \; U^- \; v_\beta 
\textrm{ .}
\end{align*}
The last two terms can be combined nicely if we note that
$\alpha = 1/\sqrt{\kappa}$ and write $x^{n-1} - y^{n-1} =
(x-y) \, (x^{n-2} + x^{n-3} y + \cdots + x y^{n-3} + y^{n-2})$.
The contribution is then
\begin{align}
\nonumber
& G_f \; \Big\{ \frac{\kappa}{2} \big(
    \alpha^2 \sum_{n,n'=1}^\infty x^{n+n'-2} a_{-n} a_{-n'}
    + \alpha \sum_{n=1}^\infty (n-1) x^{n-2} a_{-n} \big) \\
\label{eq: term 3}
& \quad + \sum_K 2 \alpha_K
      (a_{-2} + \sum_{n=3}^\infty \sum_{j=0}^{n-2} x^{j} y_K^{n-2-j} a_{-n})
    \Big\} \; U^-(x; y_1, \ldots y_M) \; v_\beta
\textrm{ .}
\end{align}

The cancellation of terms (\ref{eq: term 1}), (\ref{eq: term 2}) and
(\ref{eq: term 3}) is now a matter of direct check using
$\alpha = 1/\sqrt{\kappa}$ and $\beta = \alpha + \sum_K \alpha_K$.
In conclusion we indeed have
\begin{align*}
\sA \big( h \times G_f U^- (x; y_1, \ldots, y_M) v_\beta \big)
\; = \; 0
\end{align*}
and therefore we have proven the following.
\begin{theorem}
\label{thm: SLE kappa rho state}
For SLE${}_\kappa(\rho_1,\ldots,\rho_M)$ the ``SLE state''
\begin{align*}
M_t \; := \; & \frac{1}{Z (X_t; Y^1_t, \ldots, Y^M_t)} G_{g_t}
    V_{\alpha; \alpha_1, \ldots, \alpha_M} (X_t; Y^1_t, \ldots, Y^M_t) v_{0} \\
= \; & \const \times G_t \;
    U^-_{\alpha; \alpha_1, \ldots, \alpha_M} (X_t; Y^1_t, \ldots, Y^M_t)
    \; v_{\beta}
\textrm{ ,}
\end{align*}
where $\alpha=\frac{1}{\sqrt{\kappa}}$,
$\alpha_K = \frac{\rho_K}{2 \sqrt{\kappa}}$ and
$\beta = \alpha + \sum_K \alpha_K$,
is a $\hat{F}_\beta$ valued local martingale.
\end{theorem}

\subsection{Application to multiple SLEs}
The Coulomb gas formalism provided a convenient construction of
SLE${}_\kappa (\underline{\rho})$ state. We will see that with
screening charges it can be used to multiple SLEs as well.

The case we are interested in is that of \cite{BBK-multiple_SLEs},
which in terms of the SLE definition given in Section
\ref{sec: SLE definition} means $N \geq 2$ and $M=0$ and
$\kappa_I = \kappa \in (0,8)$ for all $I=1,\ldots,N$.

Let $0 \leq L \leq N/2$ be an integer. The screened vertex
operator
\begin{align}
\label{eq: multiple SLE vertex op}
& V^{\Gamma; L, 0}_{\alpha, \ldots, \alpha} (z_1, \ldots, z_N) \\
\nonumber
= \; & \int_\Gamma V_{\alpha, \ldots, \alpha; \alpha_-, \ldots, \alpha_-}
    (z_1, \ldots, z_N; w_1, \ldots, w_L) \;
    \ud w_1 \cdots \ud w_L
\textrm{ ,}
\end{align}
with $\alpha = \frac{1}{\sqrt{\kappa}}$ and
$\alpha_- = \frac{-2}{\sqrt{\kappa}}$, will be shown to be appropriate
for multiple SLEs. However, we postpone the discussion about the
choices of contours $\Gamma$ to Section \ref{sec: pure geometries}.

Denote
\begin{align*}
& h (z_1, \ldots, z_N; w_1, \ldots, w_L)
    = h_{\alpha, \ldots, \alpha; \alpha_-, \ldots, \alpha_-}
      (z_1,\ldots,w_L) \\
= \quad & \prod_{1 \leq I < J \leq N} (z_J- z_I)^{2/\kappa} \;
    \prod_{1 \leq R < S \leq L} (w_S - w_R)^{8/\kappa} \;
    \prod_{\substack{I=1,\ldots,N \\ R=1,\ldots,L}}
    (z_I - w_R)^{-4/\kappa}
\textrm{ ,}
\end{align*}
$\beta = N \alpha + L \alpha_- = \frac{N-2L}{\sqrt{\kappa}}$
and
\begin{align*}
Z(x_1,\ldots,x_N) \; = \; & \bra v_\beta^* ,
    V^{\Gamma; L, 0}_{\alpha, \ldots, \alpha} (x_1, \ldots, x_N)
    \; v_0 \ket \\
=\; & \int_\Gamma
    h(x_1, \ldots, x_N ; w_1, \ldots, w_L) \; \ud w_1 \cdots \ud w_L
\textrm{ .}
\end{align*}
Then $Z$ satisfies the null field equations as we will prove in the
next lemma.
\begin{lemma}
\label{lem: null field multiple SLE}
The function $Z$ defined above is annihilated by the differential
operators $\sD_I$, $I=1, \ldots, N$.
\end{lemma}
\begin{Proof}
By Lemma \ref{lem: null field CG} we have
\begin{align*}
& \sD_I h (x_1, \ldots, x_N ; w_1, \ldots, w_N) \\
= \; & \Big( \frac{\kappa}{2} \ppder{x_I} +
    \sum_{J \neq I} \big( \frac{2}{x_J-x_I} \pder{x_J}
    + \frac{(\kappa-6)/\kappa}{x_J-x_I} \big)
    \Big) \; h(x_1, \ldots, w_L) \\
= \; & \sum_{K} \big( \frac{-2}{w_K-x_I} \pder{w_K}
    + \frac{2}{w_K-x_I} \big) \; h(x_1,\ldots,w_L) \\
= \; & -2 \sum_K \pder{w_K} \Big( \frac{h(x_1,\ldots, w_L)}{w_K-x_I} \Big)
\textrm{ ,}
\end{align*}
because $h(\alpha) = \frac{6-\kappa}{2\kappa}$ and $h(\alpha_-)=1$.
Since $\Gamma$ are contours such that $h$ takes the same value
in the endpoints, this implies $\sD_I Z = 0$.
\end{Proof}
We are ready to check right away that the multiple SLE state
\begin{align*}
\frac{1}{Z(X^1_t,\ldots,x^N_t)} G_{g_t}
    V^{\Gamma; L,0}_{\alpha, \ldots, \alpha} (X^1_t,\ldots,x^N_t) v_0
\end{align*}
is a local martingale if the multiple SLE is defined by
auxiliary function (partition function) $Z$.
\begin{theorem}
\label{thm: multiple SLE state}
We have
\begin{align*}
\sA \; \Big( G_f
    \int_\Gamma \ud w_1 \cdots \ud w_L h(x_1,\ldots, w_L)
    U^-(x_1, \ldots, w_L) v_\beta \Big)
    \; = \; 0
\textrm{ .}
\end{align*}
\end{theorem}
\begin{Proof}
Again the real work was done in computations of Theorem
\ref{thm: SLE kappa rho state} and we're now just picking
the ripe fruits. The operator $\sA_I$ commutes with the
integration over $w_1,\ldots,w_L$ so we need to compute
\begin{align*}
& \sA_I \Big( h(x_1,\ldots,w_L) \times G_f U^-(x_1,\ldots,w_L) v_\beta \Big) \\
= \; & \big( \sD_I + 2 \sum_{l \leq -2} p_l (-x_I; f_{-2},\ldots)
    \pder{f_l} \big) \Big( h(x_1,\ldots,w_L) \times G_f
    U^-(x_1,\ldots,w_L) v_\beta \Big)
\textrm{ .}
\end{align*}
We first write $\sD_I h = -2 \sum_K \pder{w_K} (h / (w_K-x_I))$,
as in Lemma \ref{lem: null field multiple SLE}, and then
use Theorem \ref{thm: SLE kappa rho state} to the rest.
The result is
\begin{align*}
\; & \Big( -2 \sum_K \pder{w_K} \big(
    \frac{h(x_1, \ldots, w_L)}{w_K-x_I} \big)
    \Big) \times G_f U^-(x_1, \ldots, w_L) \; v_\beta \\
& - h(x_1, \ldots, w_L) \times \Big( \sum_K \frac{2}{w_K-x_I} \pder{w_K}
    \Big) G_f U^-(x_1, \ldots, w_L) \; v_\beta \\
= \; & -2 \sum_K \pder{w_K} \Big\{ \frac{h(x_1,\ldots,w_L)}{w_K-x_I} \times
    G_f U^-(x_1, \ldots, w_L) \; v_\beta \Big\}
\textrm{ ,}
\end{align*}
which as a total derivative vanishes after integration.
\end{Proof}
\begin{remark}
\label{rem: positivity}
Even though by Lemma \ref{lem: null field multiple SLE},
$Z$ satisfies the null field equations (b), it is not
immediately obvious that the positivity property (a) is satisfied
by $Z$ (or that $Z$ has a constant phase so that a constant multiple
of it is positive).
Unless we check this property
explicitly for our choice of contours $\Gamma$, we have no guarantee that
the driving processes (\ref{eq: driving processes}) take real values
and thus that the Loewner equation describes a growth process.
But the computations don't depend on property (a), so in this paper we
omit the question of positivity of $Z$ for multiple SLEs and focus
on the algebraic side.
\end{remark}


\section{Discussion and examples}
\label{sec: discussion}

\subsection{``Pure geometries'' and the choice of integration contours}
\label{sec: pure geometries}
\begin{figure}
\epsfbox{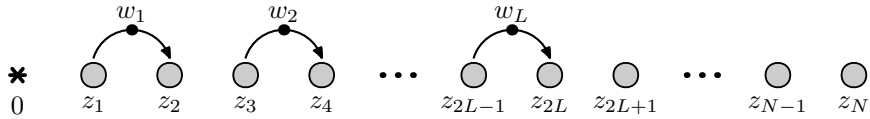}
\caption{Possible integration contours in
(\ref{eq: multiple SLE vertex op}).}
\label{fig: contour}
\end{figure}

Figure \ref{fig: contour} shows a possible integration contour for the
vertex operator (\ref{eq: multiple SLE vertex op}) of a multiple SLE.
This makes sense for $\kappa>4$, since the integrals are
convergent. But we can use the fact that
\begin{align*}
\epsfbox{VSfig.1}
\end{align*}
to continue analytically in $\kappa$ and we immediately notice that
monodromies cancel and $h$ takes the same value at the endpoints.

\begin{figure}
\includegraphics[width=1.0\textwidth]{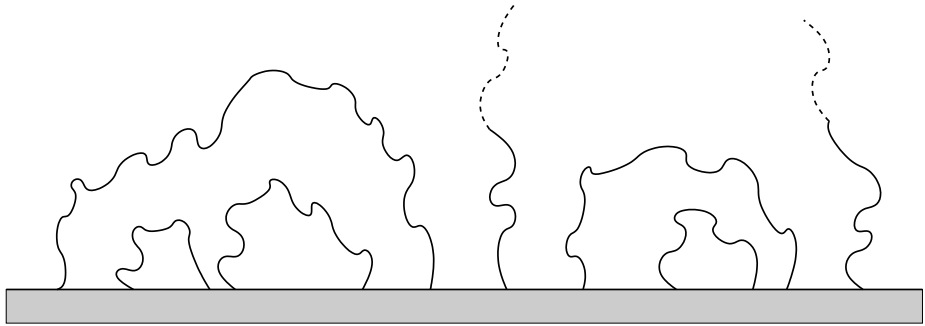}
\caption{A topological configuration of the SLE traces is conjecturally
related to the choice of integration contour for screening charges.}
\label{fig: pure geometry}
\end{figure}

The term pure geometry was introduced in \cite{BBK-multiple_SLEs}.
It was argued that to construct a multiple SLE with certain final
topological
configuration of curves one needs to require certain asymptotics
of the partition function $Z$ as some of its arguments come close
to each other. The topological configuration is the
information about how the curves are nested,
which amounts to knowing which pairs of curves will be joined.
An example configuration is illustrated in Figure \ref{fig: pure geometry}.
The choice of contours of integration of screening charges is
obviously a way of changing the asymptotics of $Z$ and
below we propose a way of getting the desired asymptotics.
Not all is proved, however. Most importantly,
the positivity of $Z$, property (a), is not obvious when
there are nested curves, see Remark \ref{rem: positivity}.

Consider $N$ simple curves in $\cl{\bH}$
starting from $z_1 < \cdots < z_N$ such that each curve either goes
to infinity without intersecting any other, or is paired with another
curve and doesn't intersect any other curve exept at the common endpoint
of its pair in $\bH$ (and this endpoint is not on any other curve!).
Let us denote the number of pairs by $L$.
Observe that the configuration is fully determined if one knows
which $z_I$ are left endpoints, meaning that the curve starting
from $z_I$ is paired with $z_J>z_I$. Namely, to reconstruct the
configuration proceed from the left. If $z_1$ is not a left endpoint
the corresponding curve goes to $\infty$. If $z_I$, $I>1$ is not
a left endpoint, there are two options. Either among $z_1,\ldots,z_{I-1}$
there are left endpoints that don't have a corresponding right endpoint
among them. In this case $z_I$ must be the right endpoint of the
righmost such left endpoint. If there are no such left endpoints,
the curve at $z_I$ must go to infinity. This way of thinking leads
to a bijection between configurations and walks
$\omega : \{ 0,1,\ldots,N \} \rightarrow \bN$ such that $\omega(0) = 0$,
$\omega(I) - \omega(I-1)=1$ if $z_I$ is not a right endpoint and
$\omega(I) - \omega(I-1)=-1$ if $z_I$ is a right endpoint. The walks
that end at $\omega(N)=N-2L$ correspond to $L$ pairs. The number of
such walks is $(N+1-2L) \frac{N!}{L! (N-L+1)!}$. Let us denote the
whole configuration by $p$.

Observe that we could have taken the integration contour in
(\ref{eq: multiple SLE vertex op}) 
to be such that integration contour of $w_R$ starts at the
$R^\textrm{th}$ left endpoint and ends at the corresponding right
endpoint (and the contours don't intersect).
Denote the corresponding
screened vertex operator by $V^{(p)}$ and partition function by
$Z^{(p)} = \bra v^*_{1,1+N-2L} , V^{(p)} v_{1,1} \ket$.
By Lemma \ref{lem: null field multiple SLE} and obvious
changes of variables we notice that $Z^{(p)}$ satisfies
(b), (c) and (d) with homogeneity degree
\begin{align*}
\Delta = L + (\frac{1}{4} N(N-1) - NL + L(L-1) ) \alpha_-^2
\textrm{ .}
\end{align*}
Since $h_{1,r} = h(\frac{1-r}{2}\alpha_-) =
\frac{1-r}{2} + \frac{r^2-1}{4}\alpha_-^2$ we see that
$\Delta = h_{1,1+N-2L} - N h_{1,2}$ in accordance with the homogeneity
degree expected of ``pure geometry'' with $N-2L$ curves going towards
infinity.

If configuration $p$ is such that $z_I$ and $z_J$ are paired,
then it is easy to see that as $|z_I-z_J| \rightarrow 0$
\begin{align*}
Z^{(p)}(z_1, \ldots, z_N) \sim (z_I - z_J)^{\frac{\kappa-6}{\kappa}}
    Z^{(p')} (z_1, \ldots, z_N)
\textrm{ ,}
\end{align*}
where $p'$ is the configuration of $N-2$ curves and $L-1$ pairs which
is obtained from $p$ by erasing the curve of $z_I$ and $z_J$.
Furthermore, considering the behavior of the integrand we expect the
asymptotic
\begin{align*}
Z^{(p)} \sim (z_I - z_J)^{\frac{2}{\kappa}}
\end{align*}
as $|z_I - z_J| \rightarrow 0$ for any two points $z_I$ and $z_J$
that are not paired.
These asymptotics are
what was in \cite{BBK-multiple_SLEs} argued for the
``pure geometry'' $p$, that is an SLE whose curves form the
configuration $p$ almost surely.

\bigskip

We still remark that the vertex operator $V^{(p)}$ with integration
contour $p$ is obtained from that of integration contour as in
Figure \ref{fig: contour}, by braid group action on the
screened vertex operators
which has been studied in \cite{FFK-braid_matrices}.

\subsection{On Möbius invariance}
\label{sec: Mobius invariance}
Consider an SLE whose state $M_t$ can be expressed as
$G_{g_t} \Psi (X^1_t, \ldots, Y^M_t) \omega_0$ and
$L_0 \omega_\infty^* = 0$ and
$\bra L_{-1} \omega_\infty^* , \Psi(x_1, \ldots, y_M) \omega_0 \ket = 0$
for all $x_1,\ldots,y_M$.
We then compute using the intertwining relation for $n=-1,0,1$
that $Z(\cdots) = \bra \omega_\infty^* ,
\Psi(\cdots) \omega_0 \ket$ satisfies the following
\begin{align*}
0 \quad = \quad & \bra \omega_{\infty}^* , \; \Psi(x_1, \ldots, y_M) \;
    L_n \; \omega_0 \ket \\
= \quad & \bra \omega_\infty^* , \; \Big( L_n \Psi - \big( \sum_{I}
    (x_I^{1+n} \pder{x_I} + (1+n) \delta_{x_I} x_I^n) \big) \Psi \\
& \qquad - \big( \sum_{K} (y_K^{1+n} \pder{y_K} + (1+n) \delta_{y_K} y_K^n)
    \big) \Psi \Big) \; \omega_0 \ket \\
= \quad & - \Big( \sum_{I} (x_I^{1+n} \pder{x_I} + (1+n) \delta_{x_I} x_I^n) \\
& \qquad + \sum_{K} (y_K^{1+n} \pder{y_K} + (1+n) \delta_{y_K} y_K^n) \Big)
    \; Z (x_1,\ldots, y_K)
\textrm{ .}
\end{align*}
The equations for $n=-1,0,1$ can be integrated to give the transformation
properties of $Z$ under translations, dilatations and special conformal
transformations
\begin{align*}
Z(x_1 - \sigma, \ldots, y_M - \sigma) \; & = \;
    Z(x_1, \ldots, y_M) \\
Z(e^{-\lambda} x_1 , \ldots, e^{-\lambda} y_M ) \; & = \;
    e^{\lambda \delta_{x_1}} \cdots e^{\lambda \delta_{y_M}} \;
    Z(x_1, \ldots, y_M) \\
Z(\frac{x_1}{1+\rho x_1}, \ldots, \frac{y_M}{1+\rho y_M} ) \; & =
    \; (1+\rho x_1)^{2 \delta_{x_1}} \cdots (1+\rho y_M)^{2 \delta_{y_M}}
    \; Z(x_1, \ldots, y_M)
\end{align*}
as long as $\rho$ is small enough so that $z \mapsto z / (1 + \rho z)$
has not mapped any of the points to $\infty$. A general Möbius
transformation $\mu : \bH \rightarrow \bH$ that preserves the order of real
points $x_1, \ldots, x_N, y_1, \ldots, y_M$ can be written as a composition
of special conformal transformation, dilatation and translation
and the transformation properties are compactly
\begin{align}
\label{eq: Moebius covariance}
Z (\mu(x_1), \ldots, \mu(y_M)) = \mu'(x_1)^{-\delta_{x_1}} \cdots
    \mu'(y_M)^{-\delta_{y_M}} \; Z (x_1, \ldots, x_N)
\textrm{ .}
\end{align}
The following Proposition says that if $Z$ is Möbius covariant in the
sense of (\ref{eq: Moebius covariance}) then the SLE variant is
Möbius invariant up to a change in growth speeds. The assertion follows
from a typical SLE computation and it can be found in
\cite{Graham-multiple_SLEs} in a slightly different form.
\begin{proposition}
Suppose the auxiliary function $Z$ of the SLE variant satisfies
(\ref{eq: Moebius covariance}) for Möbius transforms $\mu$ that
preserve the order of $x_1, \ldots, x_N, y_1, \ldots, y_M$.
Choose $\mu_t: \bH \rightarrow \bH$ Möbius such that
\begin{eqnarray*}
\tilde{g}_t = \mu_t \circ g_t \circ \mu^{-1} :
\bH \setminus \mu (K_t) \rightarrow \bH
\end{eqnarray*}
is hydrodynamically normalized. Then
$\tilde{g}_t$ describes an SLE variant with the same auxiliary
function but different growth speeds, i.e.
\begin{align*}
\ud \tilde{g}_t (z) \; = \; & \sum_I \frac{2}
    {\tilde{g}_t(z) - \tilde{X}^{I}_t} \ud \qvl \tilde{A}^{I} \qvr_t
\qquad \qquad \ud \tilde{Y}_t \; = \; \sum_I \frac{2}
    {\tilde{Y}_t - \tilde{X}^{I}_t} \ud \qvl \tilde{A}^{I} \qvr_t \\
\ud \tilde{X}^{I}_t \; = \; &
    \sqrt{\kappa} \; \ud \tilde{A}^{I}_t
    + \sum_{J \neq I} \frac{2}{\tilde{X}^{I}_t - \tilde{X}^{J}_t}
    \; \ud \qvl \tilde{A}^J \qvr_t \\
& \; + \kappa \big( \partial_{x_I} \log Z)
    (\tilde{X}^{1}_t, \ldots, \tilde{X}^{N}_t) \big)
    \; \ud \qvl \tilde{A}^I \qvr_t 
\textrm{ ,}
\end{align*}
where 
$\ud \tilde{A}^{I} = \mu_t'(X^I_t) \; \ud A^I_t$.
\end{proposition}

\begin{remark}
Examples of M\"obius invariant SLEs are e.g.
SLE${}_\kappa(\rho_1, \ldots, \rho_M)$ such that
$\sum_K \rho_K = \kappa - 6$ as in \cite{SW-coordinate_changes}
and certain pure geometries of multiple SLEs as we'll soon see.
\end{remark}

\subsection{BRST cohomology and the structure of $\sM$}
Section \ref{sec: Coulomb gas} suggests that SLE state can often be
constructed in charged Fock spaces.
To study the Fock spaces and vertex operators Felder has introduced
a method based on cohomology of BRST operators
\cite{Felder-BRST_approach}, see also
\cite{FFK-braid_matrices, BMP-Fock_space_resolutions}.
The ``BRST charge''\footnote{In the case of minimal models,
$\kappa \in \bQ$, the operators $Q$ satisfy a so called BRST property.
For $\kappa \notin \bQ$ the operators can still be defined and we
stick to the same name although it is not very meaningful.}
$Q_m$ is a $\vir$-module homomorphism $F_{n,m} \rightarrow F_{n,-m}$.
We refer the reader to \cite{Felder-BRST_approach} or \cite{DMS-CFT}
for the definition of it.
Below we show how it can be applied to questions of SLE local
martingales.

We will recall the Virasoro structure of $F_{n,m}$ in the generic
case $\kappa \notin \bQ$
and make remarks about the more complicated structure for $\kappa \in \bQ$.
Since the SLE states can in some cases be constructed in Fock spaces,
this is a step towards resolving the structure of $\sM$ because it must
be a submodule of the contravariant module.

In the generic case $\kappa \notin \bQ$ the Verma module
$\Verma_{c, h_{n,m}}$, $n,m \geq 1$, contains one singular vector $\chi_{n,m}$
that generates a submodule isomorphic to $\Verma_{c, h_{n,m}+nm}$
which is irreducible (in the classification by Feigin and Fuchs this
corresponds to case $\mathrm{II_+}$).
The kernel 
$\Kern Q_m \subset F_{n,m}$ is
a submodule and a closer study reveals that it is isomorphic to the
irreducible highest weight $\vir$-module
$\Kern Q_m \isom \irhwm_{n,m} = \Verma_{c, h_{n,m}} / \sU(\vir) \chi_{n,m}$
of highest weight $h_{n,m}$. 

A fact of great importance is that the BRST charge $Q$ commutes with
vertex operators up to a factor.
More precisely, if $V_{\underline{\alpha}}^{\Gamma; L,0}$ is a composition
of vertex operators screened with $L$ charges $\alpha_-$ such that all
$\alpha_j$'s are of the form $\alpha_{r_j, s_j}$, $r_j, s_j > 0$ and
$\sum_j \alpha_j = \alpha_{r,s}$, then
$Q_{m+s-1-2L} \; V^{\Gamma; L,0}_{\underline{\alpha}} |_{F_{n,m}} =
\const \times V^{\Gamma' ; s-L-1,0}_{\underline{\alpha}} |_{F_{n,-m}}
\; Q_{m}$,
where $\Gamma'$ is another contour of screening with $s-L-1$ charges
$\alpha_-$.
In particular, screened vertex operators map the kernel of $Q$ to the
kernel of $Q$.
Since it can be checked that
$v_0 = v_{1,1} \in \Kern Q_1$, the states of
multiple SLEs constructed in Section \ref{sec: Coulomb gas} by
Coulomb gas method take values in the submodule
$\Kern Q_{1+N-2L} \subset F_{1,1+N-2L}$.
And since $\Kern Q_{1+N-2L}$ doesn't contain the singular vector of
$F_{1,1+N-2L}$ at level $1+N-2L$ we conclude that there is a nonzero
vector at level $1+N-2L$ in $F^*_{1,1+N-2L}$ that annihilates
$\Kern Q_{1+N-2L}$. But the SLE state takes values in the annihilated
subspace so $\sM$ has a null vector at level $1+N-2L$ (for generic
$\kappa$ we readily conclude that the module $\sM$ is irreducible).
For example the null vector at level $2$ for ordinary chordal SLE
(with $N=1$, $L=0$)
can be understood in this way. A less well known case is
multiple SLE with no curves to infinity i.e. $N=2L$. There is a
vector at level $1$ in contravariant module that annihilates
$\Kern Q_1$ so the considerations of Section \ref{sec: Mobius invariance}
imply that such a multiple SLE is M\"obius invariant.

In order not to give an overly simplified picture,
let us point out that the non-generic case $\kappa \in \bQ$ is quite
involved. Write $\kappa/4 = q/q'$ such
that $q,q' \in \bN$ have no common factors.
Then we have $\alpha_{n,m} = \alpha_{n-q',m-q}$ and we can
view $Q_{q-m} : F_{n-q',q-m} \rightarrow F_{n-q',m-q} = F_{n,m}$.
The BRST property is $Q_m Q_{q-m} = 0$.
The whole structure of Fock spaces as Virasoro modules (with all
its exceptions) can be found in
\cite{FF-representations, Felder-BRST_approach,
BMP-Fock_space_resolutions}.
The charged Fock space $F_{1,m}$ contains infinitely many singular vectors
$u_i$, $i=1,2,\ldots$ and to start with one would like to know whether they
belong to the kernel of $Q_m$ or not. 
An SLE illustration of the complications is the example of
SLE${}_\kappa(\kappa-6)$ in Section \ref{sec: first examples}.
This case can be viewed as a multiple SLE (with $N=2$, $L=1$)
so that $\sM$ must contain a null vector at level $1$. But at level
$(q-1)(q'-1)$ we can have either a null vector
or a non-zero singular vector. It is worth noting that the latter
has been so far unheard of in the SLE context and it immediately rules out
the possibility of constructing the SLE state in a highest weight
module\footnote{The Fock space is of course not a $\vir$ highest weight
module.} for Virasoro algebra!


\section{Conclusions}
We have shown that local martingales for general variants of SLEs
carry a representation of the Virasoro algebra.
There exists a
natural subrepresentation $\sM$, whose interpretation was discussed.
In the general case the structure of the module of local
martingales has several properties that can not be seen in the simplest
case of the chordal SLE towards infinity \cite{BB-SLE_martingales}.
While some progress was made, the precise structure of
$\sM$ remains not completely resolved even for some of the most natural
SLE variants.

Coulomb gas method of conformal field theory was used for
constructing the SLE state explicitly in some particularly interesting
cases. From the Coulomb gas method
one obtains some results about the structure of the module as well.
In particular for multiple SLEs,
through the identification of $\sM$ as a submodule in the contravariant
module of $\Kern Q$, one gets the irreducibility of $\sM$ for
$\kappa$ generic and Möbius invariance in certain cases.
Further exploiting the BRST cohomology may be a promising approach
to a better understanding of the Virasoro structure.

The Feigin-Fuchs integrals of the Coulomb gas
give solutions to the system of differential equations needed
to define multiple SLEs and the choice of contours of screening
charges was argued to be transparently related to the conjecture
of pure geometries.

The extensive discussion of interpretation should contribute to
understanding more clearly the conformal field theoretic point of
view to SLEs. Furthermore, the sole mechanism of constructing local
martingales can turn out very useful as is illustrated for example
by a novel approach to questions of SLE duality and chordal SLE
reversibility in \cite{KK-reversibility_duality}.


\bigskip

\emph{Acknowledgements.} It is a pleasure to acknowledge that during
the writing of this article I have benefited from interesting
discussions with and useful remarks of Michel Bauer, Antti Kemppainen,
Antti Kupiainen, Luigi Cantini and Krzysztof Gaw\c edzki.
A part of this work was done
at the Department of Mathematics and Statistics at University of Helsinki
and a part at the Service de Physique Th\'eorique, CEA Saclay,
and ENRAGE European Network MRTN-CT-2004-5616.
The support of ANR-06-BLAN-0058-02 is gratefully acknowledged.

\appendix

\section{Proof of Propositions \ref{prop: Virasoro} and \ref{prop: A}}
This Appendix contains a sketch of the computations proving
Propositions \ref{prop: Virasoro} and \ref{prop: A}.
The computations are longish, but we will try to provide enough
details for a dedicated reader to follow them without too much
effort.


\subsection{Lemmas for the computations}

Certain kinds of terms will occur frequently in the computations
so we write down some lemmas for these.
\begin{lemma}
\label{lem: computation 1}
For $p \in \bN$ one has the following
\begin{align}
\nonumber
\sum_{m \leq -2} \Res{v} & v^{-2-m}
    \big( \frac{1}{f(v)-r} \big) \pder{f_m} \Big(
    \big( \frac{1}{(f(w)-s)^p}\big) \Big) \\
\label{eq: computation 1 part 1}
= \; &
    \Big( \frac{-p}{(f(w)-r) (f(w)-s)^{p+1}}
    \Big) \\
\nonumber
\sum_{m \leq -2} \Res{v} & v^{-2-m}
    \big( \frac{1}{f(v)-r} \big) \pder{f_m} 
    \big( \frac{f'(w)^2}{(f(w)-s)^p}\big) \\
\label{eq: computation 1 part 2}
= \; & f'(w)^2 \Big( \frac{-p}{(f(w)-r) (f(w)-s)^{p+1}}
    + \frac{-2}{(f(w)-r)^2 (f(w)-s)^p} \Big) \\
\nonumber
\sum_{m \leq -2} \Res{v} & v^{-2-m}
    \big( \frac{1}{f(v)-r} \big) \pder{f_m}
    \big( \frac{f'(z)^2}{(f(w)-f(z))^p}\big) \\
\nonumber
= \; & f'(z)^2 \Big( \big( \frac{1}{(f(w)-r)(f(z)-r)} \big)
    \frac{p}{(f(w)-f(z))^{p}} \\
\label{eq: computation 1 part 3}
& \qquad + \frac{-2}{(f(z)-r)^2 (f(w)-f(z))^p} \Big)
\textrm{ ,}
\end{align}
where all the rational functions are expanded in
$\Expand{f(v)}{r}$, $\Expand{f(w)}{s}$, $\Expand{f(w)}{f(z)}$ and
$\Expand{f(z)}{r}$.
\end{lemma}
\begin{Proof}
These are direct computations. First observe that
\begin{align*}
\pder{f_m} \big( \frac{1}{(f(w)-s)^p} \big)
\; = \; & \frac{-p}{(f(w)-s)^{p+1}} w^{m+1} \\
\pder{f_m} \big( \frac{f'(w)^2}{(f(w)-s)^p} \big)
\; = \; & \frac{2 f'(w)}{(f(w)-s)^p} (1+m) w^{m}
    - \frac{p f'(w)^2}{(f(w)-s)^{p+1}} w^{m+1} \\
\pder{f_m} \big( \frac{f'(z)^2}{(f(w)-f(z))^p} \big)
\; = \; & \frac{2 f'(z)}{(f(w)-f(z))^p} (1+m) z^{m} \\
& - \frac{p f'(z)^2}{(f(w)-f(z))^{p+1}} (w^{m+1} - z^{m+1})
\textrm{ .}
\end{align*}
The sums over $m$ in each case then consist of expansions of
rational functions, whose expansion we will change as follows
\begin{align*}
\sum_{m \leq -2} v^{-2-m} w^{m+1} = & \big( \frac{1}{w-v}
    \big)_{\Expand{w}{v}}
    = \big( \frac{1}{w-v} \big)_{\Expand{v}{w}} + \delta(w-v) \\
\sum_{m \leq -2} (1+m) v^{-2-m} w^{m} = & \big( \frac{-1}{(w-v)^2}
    \big)_{\Expand{w}{v}}
    = \big( \frac{-1}{(w-v)^2} \big)_{\Expand{v}{w}} - \partial_v \delta(w-v)
\end{align*}
Observe that after having changed the expansion the term involving
rational functions contains no powers of $v$ greater than $-2$ so
the residue of this part vanishes, e.g. in the case of
(\ref{eq: computation 1 part 2})
\begin{align*}
& \sum_{m \leq -2} \Res{v} v^{-2-m}
    \big( \frac{1}{f(v)-r} \big)_{\Expand{f(v)}{r}} \pder{f_m} \Big(
    f'(w)^2 \big( \frac{1}{(f(w)-s)^p}\big)_{\Expand{f(w)}{s}} \Big) \\
= \; & \Res{v} \big( \frac{1}{f(v)-r} \big)_{\Expand{f(v)}{r}}
    \Big( -p \frac{f'(w)^2}{(f(w)-s)^{p+1}} \big( \frac{1}{w-v}
        + \delta(w-v) \big) \\
& \qquad + 2 \frac{f'(w)}{(f(w)-s)^p} \big( \frac{-1}{(w-v)^2}
        - \partial_v \delta(w-v) \big)
    \Big)_{\Expand{f(w)}{s}, \; \Expand{v}{w}} \\
= \; & \Big( - p \frac{f'(w)^2}{(f(w)-r) (f(w)-s)^{p+1}}
    + \frac{2 f'(w)}{(f(w)-s)^p} \partial_w \big(\frac{1}{f(w)-r}\big)
    \Big)_{\Expand{f(w)}{r} , \; \Expand{f(w)}{s}}
\end{align*}
and the delta functions were easy to handle with an integration by
parts. Computing the remaining derivative yields the result for
(\ref{eq: computation 1 part 2}) and this in fact contains also
the computation
needed for (\ref{eq: computation 1 part 1}). For the last one,
(\ref{eq: computation 1 part 3}) we go ahead analogously
\begin{align*}
& \sum_{m \leq -2} \Res{v} v^{-2-m}
    \big( \frac{1}{f(v)-r} \big)_{\Expand{f(v)}{r}} \pder{f_m} 
    \big( \frac{f'(z)^2}{(f(w)-f(z))^p}\big)_{\Expand{f(w)}{s}}  \\
= \; & \Res{v} \big( \frac{1}{f(v)-r} \big)_{\Expand{f(v)}{r}}
    \Big( - \frac{2 f'(z)}{(f(w)-f(z))^p} \partial_v \delta(z-v) \\
& \qquad - \frac{p f'(z)^2}{(f(w)-f(z))^{p+1}}
    \big( \delta(w-v) - \delta(z-v) \big) \Big) \\
= \; & \Big( + \frac{2 f'(z)}{(f(w)-f(z))^p} \partial_z
    \big( \frac{1}{f(z)-r} \big)
    - \frac{p f'(z)^2}{(f(w)-f(z))^{p+1}}
    \big( \frac{1}{f(w)-r} - \frac{1}{f(z)-r} \big) \Big)
\textrm{ ,}
\end{align*}
where the expansions are in $\Expand{f(w)}{r}$,
$\Expand{f(z)}{r}$ and $\Expand{f(w)}{s}$.
Use $(f(w)-r)^{-1} - (f(z)-r)^{-1} = (f(z)-f(w))
\big( (f(w)-r) (f(z)-r) \big)^{-1}$ and compute the derivative
$\partial_z (f(z)-r)^{-1} = - f'(z) \, (f(z)-r)^{-2}$ to obtain
the result (\ref{eq: computation 1 part 3}).
\end{Proof}
\begin{lemma}
\label{lem: computation 2}
One has
\begin{align*}
& \sum_{m \leq -2} \Res{v} v^{-2-m}
    \big( \frac{1}{f(v)-r} \big)_{\Expand{f(v)}{r}} \pder{f_m}
    \big( Sf (w) \big) \\
= \; & - 6 f'(w)^2 \big( \frac{1}{(f(w)-r)^4} \big)_{\Expand{f(w)}{r}}
\textrm{ .}
\end{align*}
\end{lemma}
\begin{Proof}
The proof is similar to that of Lemma \ref{lem: computation 1}.
We start by computing
\begin{align*}
\pder{f_m} Sf(w) \; = \; & \pder{f_m} \big( \frac{f'''(w)}{f'(w)}
    - \frac{3}{2} \frac{f''(w)^2}{f'(w)^2} \big) \\
=\; & (m^3-m) w^{m-2} \frac{1}{f'(w)}
    - 3 (m^2+m) w^{m-1} \frac{f''(w)}{f'(w)^2} \\
& \quad + (m+1) w^{m} \big( - \frac{f'''(w)}{f'(w)^2}
        +3 \frac{f''(w)^2}{f'(w)^3} \big)
\textrm{ .}
\end{align*}
Then to change expansions we use
\begin{align*}
\sum_{m \leq -2} (m+1) v^{-2-m} w^{m} = &
    \big( \frac{-1}{(w-v)^2} \big)_{\Expand{v}{w}}
    - \partial_v \delta(w-v) \\
\sum_{m \leq -2} (m^2+m) v^{-2-m} w^{m-1} = &
    \big( \frac{2}{(w-v)^3} \big)_{\Expand{v}{w}}
    + \partial_v^2 \delta(w-v) \\
\sum_{m \leq -2} (m^3-m) v^{-2-m} w^{m-2} = &
    \big( \frac{-6}{(w-v)^4} \big)_{\Expand{v}{w}}
    - \partial_v^3 \delta(w-v)
\textrm{ .}
\end{align*}
It is already clear, just like in Lemma \ref{lem: computation 1},
that only the delta function terms will contribute because the rest
contains powers of $v$ not greater than $-2$. Thus
$\sum_m \Res{v} (f(v)-r)^{-1} v^{-2-m} \pder{f_m} Sf(w)$ can be
written as
\begin{align*}
& \Big( \big( - \frac{f'''(w)}{f'(w)^2} + 3 \frac{f''(w)^2}{f'(w)^3}
    \big) \partial_w - 3 \frac{f''(w)}{f'(w)^2} \partial_w^2
    + \frac{1}{f'(w)} \partial_w^3 \Big) \;
    \big( \frac{1}{f(w)-r} \big)_{\Expand{f(w)}{r}}
\textrm{ .}
\end{align*}
Now we substitute
\begin{align*}
\partial_w \big( \frac{1}{f(w) - r} \big) = & \frac{-f'(w)}{(f(w)-r)^2} \\
\partial_w^2 \big( \frac{1}{f(w) - r} \big) = &
    \frac{-f''(w)}{(f(w)-r)^2} + \frac{2 f'(w)^2}{(f(w)-r)^3} \\
\partial_w^3 \big( \frac{1}{f(w) - r} \big) = &
    \frac{-f'''(w)}{(f(w)-r)^2} + \frac{6 f''(w) f'(w)}{(f(w)-r)^3}
    + \frac{- 6 f'(w)^3}{(f(w)-r)^4}
\end{align*}
to obtain the asserted result.
\end{Proof}
\bigskip

\subsection{Proof of Proposition \ref{prop: Virasoro}}
\label{app: Virasoro}

We'll now get started with the computation of $[\sL_{n}, \sT(\zeta)]$.
The ``driving processes'' $x_I$ and ``passive points'' $y_K$ play the
same role in $\sT(\zeta)$ so we simplify the notation by relabeling the
$y_K$ as $x_{N+K}$. Furthermore we split
$\sT(\zeta)$ and $\sL_n$ into parts as follows
\begin{align*}
\sT(\zeta) \; = \; & \sT^{\partial f}(\zeta) + \sT^{\partial x}(\zeta)
    + \sT^{c}(\zeta) + \sT^{h}(\zeta) \qquad \textrm{, where} \\
\sT^{\partial f}(\zeta) \; = \; &
    - \sum_{l \leq -2} \zeta^{-4} f'(1/\zeta)^2 \Res{w}
    \big( \frac{w^{-2-l}}{f(w)-f(1/\zeta)} \big)_{\Expand{f(w)}{f(1/\zeta)}}
    \; \pder{f_l} \\
\sT^{\partial x}(\zeta) \; = \; &
    \sum_{K=1}^{N+M} \zeta^{-4} f'(1/\zeta)^2 \big( \frac{1}{f(1/\zeta)- x_K}
    \big)_{\Expand{f(1/\zeta)}{x_K}} \; \pder{x_K} \\
\sT^{c}(\zeta) \; = \; &
    \frac{c}{12} \zeta^{-4} Sf(1/\zeta) \\
\sT^{h}(\zeta) \; = \; &
    \sum_{K=1}^{N+M} \zeta^{-4} f'(1/\zeta)^2 \big( \frac{\delta_{x_K}}
    {(f(1/\zeta)-x_K)^2} \big)_{\Expand{f(1/\zeta)}{x_K}}
\textrm{ ,}
\end{align*}
and similarly
$\sL_n = \Res{u} u^{-3-n} \sT(u^{-1}) =
\sL^{\partial f}_n + \sL^{\partial y}_n + \sL^c_n + \sL^h_n$.

We will do the computation in four steps, gradually working
through special cases and finally achieving the full result.
The intermediate results can sometimes be very useful, too.

\subsubsection*{Step I: $\sT^{\partial f}$}
Let us first compute the part of the commutator in the simplest case
$c=0$, $N=M=0$, in which $\sT$ contains only the $\sT^{\partial f}$ term.
We compute
\begin{align*}
& [\sL^{\partial f}_n , \sT^{\partial f}] \\
= \; &
    \sum_{l \leq -2} \Res{z} \Res{w} \Res{u} \Big\{
    \frac{u^{1-n} f'(u)^2}{f(z)-f(u)} \sum_{k \leq -2} z^{-2-k} \pder{f_k}
    \big( \zeta^{-4} \frac{f'(1/\zeta)^2 w^{-2-l}}{f(w)-f(1/\zeta)} \big) \\
& \quad - \zeta^{-4} \frac{f'(1/\zeta)^2}{f(w)-f(1/\zeta)} \sum_{k \leq -2}
    w^{-2-k} \pder{f_k} \big( \frac{u^{1-n} f'(u)^2 z^{-2-l}}{f(z)-f(u)}
    \big) \Big\} \; \pder{f_l}
\textrm{ ,}
\end{align*}
where the expansions of rational functions are in
$|f(w)| > |f(1/\zeta)|$, $|f(z)|>|f(u)|$.
We now apply to both terms Lemma \ref{lem: computation 1}
(\ref{eq: computation 1 part 3}) to write this as
\begin{align*}
& \sum_{l \leq -2} \Res{u} \Big\{ \Res{w}  w^{-2-l} \Big(
    \frac{ u^{1-n} f'(u)^2 \zeta^{-4} f'(1/\zeta)^2
        \big( f(1/\zeta) + f(u) - 2 f(w) \big)}
        {(f(w)-f(1/\zeta)) (f(w)-f(u)) (f(1/\zeta)-f(u))^2}
    \Big)_{\Expand{f(1/\zeta)}{f(u)}} \\
& - \Res{z} z^{-2-l} \frac{ u^{1-n} f'(u)^2 \zeta^{-4} f'(1/\zeta)^2
        \big( f(1/\zeta) + f(u) - 2 f(z) \big)}
        {(f(z)-f(1/\zeta)) (f(z)-f(u)) (f(1/\zeta)-f(u))^2}
    \Big)_{\Expand{f(u)}{f(1/\zeta)}} \Big\} \pder{f_l}
\textrm{ .}
\end{align*}
We rename the dummy variable $z$ as $w$ and observe that
the rational functions would cancel if they were expanded in
the same region.
So we will change the expansion of the former. To do this,
we note that (\ref{eq: change of variables}) can be used to
show
\begin{align*}
\big( \frac{f'(\rho)}{f(u)-f(\rho)} \big)_{\Expand{f(u)}{f(\rho)}}
- \big( \frac{f'(\rho)}{f(u)-f(\rho)} \big)_{\Expand{f(\rho)}{f(u)}}
= \delta(u-\rho)
\textrm{ .}
\end{align*}
With a substitution $\rho = 1/\zeta$ and a derivative w.r.t. $u$
we obtain
\begin{align*}
- \big( \frac{f'(u) f'(\rho)}{(f(u)-f(1/\zeta))^2}
    \big)_{\Expand{f(u)}{f(1/\zeta)}}
& + \big( \frac{f'(u) f'(\rho)}{(f(u)-f(1/\zeta))^2}
    \big)_{\Expand{f(1/\zeta)}{f(u)}} \\
\; = \; &  \partial_u \delta(u-1/\zeta)
\textrm{ .}
\end{align*}
Taking the residue will therefore be easily treated as soon as we have
computed
\begin{align*}
& \partial_u \Big( \frac{ u^{1-n} f'(u) 
        \big( f(1/\zeta) + f(u) - 2 f(w) \big)}
        {(f(w)-f(u))} \Big) \Big|_{u=1/\zeta} \\
= \; & \Big( - 2 \big( (1-n) \zeta^{n} f'(1/\zeta) + \zeta^{n-1}
        f''(1/\zeta) \big)
    - \zeta^{n-1} \frac{ f'(1/\zeta)^2}{(f(w)-f(1/\zeta))} \Big)
\end{align*}
which is needed when integrating by parts. The result for
$[\sL^{\partial f}_n , \sT^{\partial f} (\zeta)]$ is
\begin{align*}
\cdots \; = \; & - \sum_{l \leq -2} \Res{w} w^{-2-l}
    \big(\frac{\zeta^{-4} f'(1/\zeta)}{(f(w)-f(1/\zeta))} \big)
    \Big( - \zeta^{n-1} \frac{ f'(1/\zeta)^2}{(f(w)-f(1/\zeta))} \\
& \qquad - 2 \big( (1-n) \zeta^{n} f'(1/\zeta) + \zeta^{n-1} f''(1/\zeta)
    \big) \Big) \; \pder{f_l} \\
= \; & \sum_{l \leq -2} \Res{w} w^{-2-l}
    \big(\frac{1}{(f(w)-f(1/\zeta))} \big)
    \Big( 2 (1-n) \zeta^{n-4} \frac{f'(1/\zeta)^2}{(f(w)-f(1/\zeta))^2} \\
& \quad + 2 \zeta^{n-5} \frac{f''(1/\zeta) f'(1/\zeta)}{(f(w)-f(1/\zeta))^2}
    + \zeta^{n-5} \frac{ f'(1/\zeta)^3}{(f(w)-f(1/\zeta))^2}
    \Big) \; \pder{f_l}
\textrm{ .}
\end{align*}
But when we compare this and
\begin{align*}
& \partial_\zeta \big( - \zeta^{-4}
    \frac{f'(1/\zeta)^2}{f(w)-f(1/\zeta)} \big) \\
= \; & 4 \zeta^{-5} \frac{f'(1/\zeta)^2)}{f(w)-f(1/\zeta)}
    + 2 \zeta^{-6} \frac{f''(1/\zeta) f'(1/\zeta)}{f(w)-f(1/\zeta)}
    + \zeta^{-6} \frac{f'(1/\zeta)^3}{(f(w)-f(1/\zeta))^2}
\end{align*}
we get the desired result
\begin{align*}
[ \sL^{\partial f}_n , \sT^{\partial f} (\zeta)] \; = \;
    (1+n) \zeta^n \big( 2 \sT^{\partial f} (\zeta) \big)
    + \zeta^{1+n} \big( \partial_\zeta \sT^{\partial f} (\zeta) \big)
\textrm{ ,}
\end{align*}
which also means that the operators $\sL^{\partial f}_n$ satisfy
the Witt algebra
\begin{align*}
[\sL^{\partial f}_n , \sL^{\partial f}_m]
    = (n-m) \sL^{\partial f}_{n+m}
\textrm{ .}
\end{align*}

\subsubsection*{Step II: $\sT^{\partial f} + \sT^c$}
Next consider terms involving the central charge $c$ in the
commutator $[\sL_n , \sT(\zeta)]$. These are
\begin{align*}
& \sL^{\partial f}_n \sT^c(\zeta) - \sT^{\partial f} \sL^c_n \\
= \; & - \frac{c}{12} \Res{u} \Res{z} \sum_{l \leq -2} \Big(
    \frac{u^{1-n} f'(u)^2 z^{-2-l}}{f(z)-f(u)} \;
    \frac{\partial}{\partial f_l} Sf(1/\zeta) \zeta^{-4} \\
& \qquad - \frac{\zeta^{-4} f'(1/\zeta)^2 z^{-2-l}}{f(z)-f(1/\zeta)}
    \; \frac{\partial}{\partial f_l} Sf(u) u^{1-n}
    \Big)
\textrm{ .}
\end{align*}
We apply Lemma \ref{lem: computation 2} to both terms and the above
simplifies to
\begin{align*}
&\frac{c}{2} \; \Res{u} u^{1-n} f'(u)^2 \zeta^{-4} f'(1/\zeta)^2
    \Big( \big( \frac{1}{(f(1/\zeta)-f(u))^4}
        \big)_{\Expand{f(1/\zeta)}{f(u)}} \\
& \qquad    - \big( \frac{1}{(f(1/\zeta)-f(u))^4}
        \big)_{\Expand{f(1/\zeta)}{f(u)}} \Big)
\textrm{ .}
\end{align*}
Thus we only have the
contribution from changing the expansions,
the ``residue at $u=1/\zeta$''. The straightforward evaluation of
this gives
\begin{align*}
& \frac{c}{12} \Big( \zeta^{n-2} (n^3-n) - 3 \zeta^{n-5}
    \frac{f''(1/\zeta)^3}{f'(1/\zeta)^3}
    - 3 (n-1) \zeta^{n-4} \frac{f''(1/\zeta)^2}{f'(1/\zeta)^2} \\
& + 4 \zeta^{n-5} \frac{f''(1/\zeta) f'''(1/\zeta)}{f'(1/\zeta)^2}
    + 2 (n-1) \zeta^{n-4} \frac{f'''(1/\zeta)}{f'(1/\zeta)}
    - \zeta^{n-5} \frac{f''''(1/\zeta)}{f'(1/\zeta)} \Big) \\
= \; & \frac{c}{12} \Big( \zeta^{n-2} (n^3-n)
    + (1+n) \zeta^{n} \; \big( 2 \zeta^{-4} Sf(1/\zeta) \big)
    + \zeta^{1+n} \; \partial_\zeta \big( \zeta^{-4} Sf(1/\zeta) \big) \Big)
\textrm{ ,}
\end{align*}
as it should. In other words, we have
\begin{align*}
& [\sL^{\partial f}_n + \sL^{c}_n ,
    \sT^{\partial f}(\zeta) + \sT^{c}(\zeta)] \\
= \; & \frac{c}{12} \zeta^{n-2} (n^3-n) + (1+n) \zeta^{n} \;
      2 \big( \sT^{\partial f} (\zeta) + \sT^c(\zeta) \big)
    + \zeta^{1+n} \; \partial_\zeta
    \big( \sT^{\partial f} (\zeta) + \sT^c(\zeta) \big)
\end{align*}
and thus $\sL^{\partial f}_n + \sL^{c}_n$ satisfy the Virasoro algebra
\begin{align*}
& [ ( \sL^{\partial f}_n + \sL^{c}_n ) , ( \sL^{\partial f}_m + \sL^{c}_m )] \\
= \; & (n-m) ( \sL^{\partial f}_{n+m} + \sL^{c}_{n+m} )
    + \frac{c}{12} (n^3-n) \delta_{n+m,0}
\textrm{ .}
\end{align*}

\subsubsection*{Step III:
$\sT^{\partial f} + \sT^{c} + \sT^{\partial y}$}
The remaining task is to compute the terms involving the
points $y_K$ and we will start with the special case of vanishing
$\delta_{x_I}$.

Let us first aim at finding out what is
$\sL^{\partial x}_n + \sL^{\partial f}_n$
acting on $\sT^{\partial x}$. We use Lemma \ref{lem: computation 1}
(\ref{eq: computation 1 part 2}) to get
\begin{align*}
& \Res{u} u^{1-n} f'(u)^2 \Big( \sum_K \frac{1}{f(u)-x_K} \pder{x_K} \\
& \qquad - \Res{z} \sum_{l \leq -2} \frac{z^{-2-l}}{f(z)-f(u)} \pder{f_l}
    \Big) \; \frac{\zeta^{-4} f'(1/\zeta)^2}{f(1/\zeta) - x_J} \\
= \; & \Res{u} u^{1-n} f'(u)^2 \zeta^{-4} f'(1/\zeta)^2
    \Big( \frac{1}{(f(1/\zeta)-f(u)) (f(1/\zeta)-x_J)^2} \\
& \qquad + \frac{2}{(f(1/\zeta)-f(u))^2 (f(1/\zeta)-x_J)}
    + \frac{1}{(f(u)-x_J)(f(1/\zeta)-x_J)^2} \Big) \\
= \; & \Res{u} u^{1-n} f'(u)^2 \zeta^{-4} f'(1/\zeta)^2
    \Big( \frac{f(u) + f(1/\zeta) - 2 x_J}
    {(f(1/\zeta)-f(u))^2 (f(u)-x_J) (f(1/\zeta)-x_J)} \Big)
\end{align*}
expanded in $\Expand{f(1/\zeta)}{f(u)}$. One does a completely
analogous computation for
$\sT^{\partial x} + \sT^{\partial f}$ acting on
$\sL^{\partial x}_n$ to yield a nice looking result
\begin{align*}
& \big( \sL^{\partial x}_n + \sL^{\partial f}_n \big)
        \sT^{\partial x}(\zeta)
    - \big( \sT^{\partial x} (\zeta) + \sT^{\partial f} (\zeta) \big)
        \sL^{\partial x}_n \\
= \; & \sum_J \Res{u} u^{1-n} f'(u)^2 \zeta^{-4} f'(1/\zeta)^2
    \Big( \frac{f(u) + f(1/\zeta) - 2 x_J}
    {(f(u)-x_J) (f(1/\zeta)-x_J)} \Big) \\
& \qquad \Big(
    \big( \frac{1}{(f(1/\zeta)-f(u))^2} \big)_{\Expand{f(1/\zeta)}{f(u)}}
    - \big( \frac{1}{(f(1/\zeta)-f(u))^2} \big)_{\Expand{f(u)}{f(1/\zeta)}}
    \Big) \; \pder{x_J}
\textrm{ .}
\end{align*}
We're again in a position to change the expansions to get the
``residue at $u = 1/\zeta$''. The formula for change of expansion
was given in Step I. The integration by parts now essentially
consists of computing
\begin{align*}
& \partial_u \Big( \frac{u^{1-n} f'(u) \big( f(u) + f(1/\zeta) - 2 x_J \big)}
    {(f(u)-x_J)} \Big) \Big|_{u = 1/\zeta} \\
= \; & \Big( 2 \big( (1-n) \zeta^{n} f'(1/\zeta)
        + \zeta^{n-1} f''(1/\zeta) \big)
    - \zeta^{n-1} \frac{f'(1/\zeta)^2}{(f(1/\zeta)-x_J)} \Big)
\textrm{ .}
\end{align*}
Now we can write
\begin{align*}
& \big( \sL^{\partial x}_n + \sL^{\partial f}_n \big)
        \sT^{\partial x}(\zeta)
    - \big( \sT^{\partial x} (\zeta) + \sT^{\partial f} (\zeta) \big)
        \sL^{\partial x}_n \\
= \; & - \sum_J \Big( \frac{\zeta^{-4} f'(1/\zeta)}{(f(1/\zeta)-x_J)} \Big)
    \Big( 2 \big( (1-n) \zeta^{n} f'(1/\zeta)
        + \zeta^{n-1} f''(1/\zeta) \big) \\
& \qquad - \zeta^{n-1} \frac{f'(1/\zeta)^2}{(f(1/\zeta)-x_J)}
    \Big) \; \pder{x_J} \\
= \; & \sum_J \Big( - 2 (1-n) \zeta^{n-4} \frac{f'(1/\zeta)^2}{f(1/\zeta)-x_J}
        -2 \zeta^{n-5} \frac{f''(1/\zeta) f'(1/\zeta)}{f(1/\zeta)-x_J}  \\
& \qquad + \zeta^{n-5} \frac{f'(1/\zeta)^3}{(f(1/\zeta)-x_J)^2}
    \Big) \; \pder{x_J}
\end{align*}
and compare with
\begin{align*}
& \partial_\zeta \Big( \zeta^{-4} \frac{f'(1/\zeta)^2}{f(1/\zeta) - x_J}
    \Big) \\
= \; & -4 \zeta^{-5} \frac{f'(1/\zeta)^2}{f(1/\zeta) - x_J}
    - 2 \zeta^{-6} \frac{f''(1/\zeta) f'(1/\zeta)}{f(1/\zeta) - x_J}
    + \zeta^{-6} \frac{f'(1/\zeta)^3}{(f(1/\zeta) - x_J)^2}
\end{align*}
to notice that
\begin{align*}
& \big( \sL^{\partial x}_n + \sL^{\partial f}_n \big)
        \sT^{\partial x}(\zeta)
    - \big( \sT^{\partial x} (\zeta) + \sT^{\partial f} (\zeta) \big)
        \sL^{\partial x}_n \\
= \; & (1+n) \zeta^{n} \big( 2 \sT^{\partial x} (\zeta) \big)
    + \zeta^{1+n} \big( \partial_\zeta \sT^{\partial x} (\zeta) \big)
\textrm{ .}
\end{align*}
Combining with Step I and Step II we have shown that
\begin{align*}
& [\sL^{\partial f}_n + \sL^{c}_n + \sL^{\partial x}_n,
    \sT^{\partial f}(\zeta) + \sT^{c}(\zeta) + \sT^{\partial x}(\zeta)] \\
= \; & \frac{c}{12} \zeta^{n-2} (n^3-n) + (1+n) \zeta^{n} \;
      2 \big( \sT^{\partial f} (\zeta) + \sT^c(\zeta)
    + \sT^{\partial x}(\zeta) \big) \\
& + \zeta^{1+n} \; \partial_\zeta
    \big( \sT^{\partial f} (\zeta) + \sT^c(\zeta)
    + \sT^{\partial x}(\zeta) \big)
\end{align*}
and again $\sL^{\partial f}_n + \sL^{c}_n + \sL^{\partial y}_n$
satisfy the Virasoro algebra
\begin{align*}
& [ ( \sL^{\partial f}_n + \sL^{c}_n + \sL^{\partial x}_n) ,
    ( \sL^{\partial f}_m + \sL^{c}_m + \sL^{\partial x}_m)] \\
= \; & (n-m) ( \sL^{\partial f}_{n+m} + \sL^{c}_{n+m}
    + \sL^{\partial x}_{n+m} )
    + \frac{c}{12} (n^3-n) \delta_{n+m,0}
\textrm{ .}
\end{align*}

\subsubsection*{Step IV:
$\sT^{\partial f} + \sT^c + \sT^{\partial y} + \sT^{h}$}
The last piece to take into account is $\sT^{h}$, that is to allow
nonvanishing $\delta_{x_K}$. Its treatment is very similar to
Step III. We'd like to compute what is
$\sL^{\partial x}_n + \sL^{\partial f}_n$ acting on $\sT^{h}$
so we begin by using Lemma \ref{lem: computation 1}
(\ref{eq: computation 1 part 2})
\begin{align*}
& \Res{u} u^{1-n} f'(u)^2 \Big(
    \sum_K \frac{1}{f(u)-x_K} \pder{x_K} \\
& \qquad - \Res{z} \sum_{l \leq -2} \frac{z^{-2-l}}{f(z)-f(u)} \pder{f_l}
    \Big) \; \frac{\delta_{x_J} \, \zeta^{-4} f'(1/\zeta)^2}
    {(f(1/\zeta)-x_J)^2} \\
= \; & \Res{u} u^{1-n} f'(u)^2 \zeta^{-4} f'(1/\zeta)^2
    \Big( \frac{2 \delta_{x_J}}{(f(1/\zeta)-f(u)) (f(1/\zeta)-x_J)^3} \\
& \qquad + \frac{2 \delta_{x_J}}{(f(1/\zeta)-f(u))^2 (f(1/\zeta)-x_J)^2}
    + \frac{2 \delta_{x_J}}{(f(u)-x_J) (f(1/\zeta)-x_J)^3} \Big) \\
= \; & \Res{u} u^{1-n} f'(u)^2 \zeta^{-4} f'(1/\zeta)^2
    \Big( \frac{ 2 \delta_{x_J}}
    {(f(1/\zeta)-f(u))^2 (f(1/\zeta)-x_J) (f(u)-x_J)} \Big)
\end{align*}
expanded in $\Expand{f(1/\zeta)}{f(u)}$. Do the same for
$\sT^{\partial x} + \sT^{\partial f}$ acting on $\sL^{h}_n$
and obtain
\begin{align*}
& \big( \sL^{\partial x}_n + \sL^{\partial f}_n \big) \sT^{h}(\zeta)
- \big( \sT^{\partial x}(\zeta) + \sT^{\partial f}(\zeta) \big) \sL^{h}_n \\
= \; & \sum_J \Res{u} u^{1-n} f'(u)^2 \zeta^{-4} f'(1/\zeta)^2
    \Big( \frac{ 2 \delta_{x_J}}
    {(f(1/\zeta)-x_J) (f(u)-x_J)} \Big) \\
& \qquad \Big( \big( \frac{1}{(f(1/\zeta)-f(u))^2}
        \big)_{\Expand{(f(1/\zeta)}{(f(u)}}
    - \big( \frac{1}{(f(1/\zeta)-f(u))^2}
        \big)_{\Expand{f(u)}{(f(1/\zeta)}} \Big)
\textrm{ .}
\end{align*}
After changing the expansion with the formula found in Step I
we need the following
\begin{align*}
& \partial_u \Big( \frac{ u^{1-n} f'(u)}
    {(f(u)-x_J)}  \Big) \Big|_{u = 1/\zeta} \\
= \; & (1-n) \zeta^{n} \frac{f'(1/\zeta)}{(f(u)-x_J)}
    + \zeta^{n-1} \frac{f''(1/\zeta)}{(f(u)-x_J)}
    - \zeta^{n-1} \frac{f'(1/\zeta)^2}{(f(u)-x_J)^2}
\end{align*}
for integration by parts. The result can now be written as
\begin{align*}
& \big( \sL^{\partial x}_n + \sL^{\partial f}_n \big) \sT^{h}(\zeta)
- \big( \sT^{\partial x}(\zeta) + \sT^{\partial f}(\zeta) \big) \sL^{h}_n \\
= \; & - \sum_J \frac{2 \delta_{x_J} \zeta^{-4} f'(1/\zeta)}{(f(1/\zeta)-x_J)}
    \Big( \frac{(1-n) \zeta^{n} f'(1/\zeta)}{(f(u)-x_J)}
    + \frac{\zeta^{n-1} f''(1/\zeta)}{(f(u)-x_J)}
    - \frac{\zeta^{n-1} f'(1/\zeta)^2}{(f(u)-x_J)^2} \Big) \\
= \; & \sum_J 2 \delta_{x_J}
    \Big( - \frac{(1-n) \zeta^{n-4} f'(1/\zeta)^2}{(f(u)-x_J)^2}
    - \frac{\zeta^{n-5} f''(1/\zeta) f'(1/\zeta)}{(f(u)-x_J)^2}
    + \frac{\zeta^{n-5} f'(1/\zeta)^3}{(f(u)-x_J)^3} \Big)
\end{align*}
and comparison with
\begin{align*}
& \partial_{\zeta} \Big( \zeta^{-4}
    \frac{f'(1/\zeta)^2}{(f(1/\zeta)-x_J)^2} \Big) \\
= \; & -4 \zeta^{-5} \frac{f'(1/\zeta)^2}{(f(1/\zeta)-x_J)^2}
    - 2 \zeta^{-6} \frac{f''(1/\zeta) f'(1/\zeta)}{(f(1/\zeta)-x_J)^2}
    + 2 \zeta^{-6} \frac{f'(1/\zeta)^3}{(f(1/\zeta)-x_J)^3}
\end{align*}
yields the anticipated result
\begin{align*}
& \big( \sL^{\partial x}_n + \sL^{\partial f}_n \big) \sT^{h}(\zeta)
- \big( \sT^{\partial x}(\zeta) + \sT^{\partial f}(\zeta) \big) \sL^{h}_n \\
= \; & (1+n) \zeta^{n} \big( 2 \sT^h (\zeta) \big)
    + \zeta^{1+n} \big( \partial_\zeta \sT^h (\zeta) \big)
\textrm{ .}
\end{align*}

This concludes the proof of Proposition \ref{prop: Virasoro} since
together with results of Steps I-III it means
\begin{align*}
[\sL_n , \sT (\zeta)] \; = \; (1+n) \zeta^n \big( 2 \sT (\zeta) \big)
    + \zeta^{1+n} \big( \partial_\zeta \sT (\zeta) \big)
\end{align*}
which in turn is equivalent to the $\sL_n$ forming a representation
of the Virasoro algebra
\begin{align*}
[\sL_n , \sL_m] \; = \; (n-m) \sL_{n+m}
    + \frac{c}{12} (n^3-n) \delta_{n+m,0}
\textrm{ . } \quad \square 
\end{align*}


\subsection{Proof of Proposition \ref{prop: A}}
\label{app: T commutator A}
This appendix contains the computations proving
Proposition \ref{prop: A}.
We will continue to denote $y_K$ by $x_{N+K}$ for simplicity
of notation.
Thus we will use for $I=1,\ldots,N$ (for the variables corresponding
to the actual driving processes)
\begin{eqnarray*}
\sA_I & = & \frac{\kappa_I}{2} \ppder{x_I} + \sum_{J \neq I} \big(
    \frac{2}{x_J-x_I} \pder{x_J} - \frac{2 h_{x_J}}{(x_J-x_I)^2} \big) \\
& & + 2 \sum_{m \leq -2} \Res{v} v^{-2-m} \big( \frac{1}{f(v)-x_I}
    \big)_{\expand{f(v)}{large}} \; \pder{f_m}
\textrm{ ,}
\end{eqnarray*}
where the first line corresponds to the operator $\sD_I$ of null
field equation (b).
One should observe that
for the statement about $[\sT(\zeta), \sA_I]$ it will only be
necessary to require $c=\frac{(3\kappa_I-8)(6-\kappa_I)}{2\kappa_I}$
and $\delta_{x_I} = h_{x_I} = \frac{6-\kappa_I}{2 \kappa_I}$
but $\delta_{x_J} = h_{x_J}$ for $J \neq I$ can take any values
(in particular this is the case of the passive points $y_K$ now
labeled as $x_{N+K}$).

Let's start by applying Lemma \ref{lem: computation 1}
(\ref{eq: computation 1 part 1}) \& (\ref{eq: computation 1 part 3})
to the part
\begin{align*}
& [\sT^{\partial f} (\zeta) ,
    2 \, \sum_{m \leq -2} \Res{v} \frac{v^{-2-m}}{f(v)-x_I} \pder{f_m}] \\
= \; & - \sum_{l \leq -2} \Res{w} \zeta^{-4} 
    \frac{f'(1/\zeta)^2 w^{-2-l}}{f(w)-f(1/\zeta)} \sum_{m \leq -2}
    \pder{f_l} \Big( \Res{v} \frac{2 v^{-2-m}}{f(v)-x_I} \Big) \pder{f_m} \\
&   - \sum_{m \leq -2} \Res{v} \frac{2 v^{-2-m}}{f(v)-x_I}
    \sum_{l \leq -2} \pder{f_m} \Big( - \zeta^{-4}
    \frac{f'(1/\zeta)^2 w^{-2-l}}{f(w)-f(1/\zeta)} \Big) \pder{f_l} \\
= \; & 2 \sum_{m \leq -2} \Res{v} \zeta^{-4}
    \frac{f'(1/\zeta)^2}{f(v)-f(1/\zeta)} \frac{v^{-2-m}}{(f(v)-x_I)^2}
    \; \pder{f_m} \\
& + 2 \sum_{l \leq -2} \Res{w} \zeta^{-4} 
    \frac{f'(1/\zeta)^2 w^{-2-l}}{f(w)-f(1/\zeta)}
    \Big( \frac{1}{(f(w)-x_I) (f(1/\zeta)-x_I)} \\
& \qquad + \frac{-2}{(f(1/\zeta)-x_I)^2} \Big) \; \pder{f_l}
\textrm{ .}
\end{align*}
We'll rename the dummy variables $w$ as $v$ and $l$ as $m$ to
combine the two terms. In addition we take into account
the term $\sT^{\partial x}$ acting on
$\sum_{m} \Res{v} \frac{2 v^{-2-m}}{f(v)-x_I} \pder{f_m}$,
that is
\begin{align*}
2 \sum_{m \leq -2} \Res{v} \zeta^{-4}
    \frac{f'(1/\zeta)^2}{f(1/\zeta) - x_I} \frac{v^{-2-m}}{(f(v)-x_I)^2}
    \; \pder{f_m}
\textrm{ .}
\end{align*}
The sum of the terms considered above is
\begin{align}
\label{eq: T comm A part 1}
-4 \, \zeta^{-4} \frac{f'(1/\zeta)^2}{(f(1/\zeta)-x_I)^2}
\sum_{m \leq -2} \Res{v} \frac{v^{-2-m}}{f(v)-x_I} \pder{f_m}
\textrm{ .}
\end{align}

The only term that will contain the central charge $c$ comes
from the action of
$\sum_{m} \Res{v} \frac{2 v^{-2-m}}{f(v)-x_I} \pder{f_m}$
on $\sT^c$, and it is readily computed with the help of
Lemma \ref{lem: computation 2}
\begin{align}
\nonumber
& [ \sT^c (\zeta) , 2 \, \sum_{m \leq -2} \Res{v} \frac{v^{-2-m}}{f(v)-x_I}
    \pder{f_m} ] \\
\nonumber
= \; & -2 \, \sum_{m \leq -2} \Res{v} \frac{v^{-2-m}}{f(v)-x_I}
    \pder{f_m} \Big( \frac{c}{12} \zeta^{-4} Sf (1/\zeta) \Big) \\
\label{eq: T comm A part 2}
= \; & c \; \zeta^{-4} \frac{f'(1/\zeta)^2}{(f(1/\zeta)-x_I)^4}
\textrm{ .}
\end{align}

Next, let's see what are the contributions of
$\sum_{m} \Res{v} \frac{2 v^{-2-m}}{f(v)-x_I} \pder{f_m}$
acting on $\sT^{\partial x}$ and $\sT^{h}$. In both cases
Lemma \ref{lem: computation 1} (\ref{eq: computation 1 part 2})
will be used.
The former is
\begin{align*}
& -2 \, \sum_{m \leq -2} \Res{v} \frac{v^{-2-m}}{f(v)-x_I}
    \sum_J \pder{f_m} \Big( \zeta^{-4}
    \frac{f'(1/\zeta)^2}{f(1/\zeta)-x_J} \Big) \; \pder{x_J} \\
= \; & \sum_J \Big(
    \frac{2 \zeta^{-4} f'(1/\zeta)^2}{(f(1/\zeta)-x_I) (f(1/\zeta)-x_J)^2}
    + \frac{4 \zeta^{-4} f'(1/\zeta)^2}{(f(1/\zeta)-x_I)^2 (f(1/\zeta)-x_J)}
    \Big) \; \pder{x_J}
\end{align*}
and the latter
\begin{align*}
& -2 \, \sum_{m \leq -2} \Res{v} \frac{v^{-2-m}}{f(v)-x_I}
    \sum_J \pder{f_m} \Big( \zeta^{-4}
    \frac{\delta_{x_J} f'(1/\zeta)^2}{f(1/\zeta)-x_J} \Big) \\
= \; & \sum_J \Big(
    \frac{4 \delta_{x_J} \zeta^{-4} f'(1/\zeta)^2}
        {(f(1/\zeta)-x_I) (f(1/\zeta)-x_J)^3}
    + \frac{4 \delta_{x_J} \zeta^{-4} f'(1/\zeta)^2}
        {(f(1/\zeta)-x_I)^2 (f(1/\zeta)-x_J)^2} \Big)
\textrm{ .}
\end{align*}
What remains is the commutator of $\sT$ with $\sD_I$. This is
\begin{align*}
& [\sT^{\partial y} (\zeta) + \sT^{h} (\zeta) \, , \, \sD_I ] \\
= \; & \zeta^{-4} f'(1/\zeta)^2 \Big\{
    - \frac{\kappa_I}{2} \frac{2}{(f(1/\zeta)-x_I)^2} \ppder{x_I}
    - \frac{\kappa_I}{2} \frac{6 \delta_{x_I}}{(f(1/\zeta)-x_I)^4} \\
& \quad - \kappa_I \big( \frac{1}{(f(1/\zeta)-x_I)^3}
        + \frac{2 \delta_{x_I}}{(f(1/\zeta)-x_I)^3} \big) \pder{x_I} \\
& \quad - \sum_{J \neq I} \frac{2}{x_J-x_I} \Big(
        \frac{1}{(f(1/\zeta)-x_J)^2} \pder{x_J}
        + \frac{2 \delta_{x_J}}{(f(1/\zeta)-x_J)^3} \Big) \\
& \quad + \sum_{J \neq I} \frac{x_J-x_I}{(f(1/\zeta)-x_J) (f(1/\zeta)-x_I)}
        \Big( \frac{-2}{(x_J-x_I)^2} \pder{x_J}
        + \frac{4 h_{x_J}}{(x_J-x_I)^3} \Big)
    \Big\}
\textrm{ .}
\end{align*}
We will collect the different terms from $[\sT, \sD_I]$ and
$\sum_{m} \Res{v} \frac{2 v^{-2-m}}{f(v)-x_I} \pder{f_m}$
acting on $\sT^{\partial x}$ and $\sT^{h}$. The terms with
derivatives with respect to $x_I$ are
\begin{align}
\label{eq: T comm A part 3}
- \kappa_I \; \zeta^{-4} \frac{f'(1/\zeta)^2}{(f(1/\zeta)-x_I)^2} \ppder{x_I}
+  \zeta^{-4} \frac{f'(1/\zeta)^2 (6 -\kappa_I - 2 \kappa_I \delta_{x_I})}
    {(f(1/\zeta)-x_I)^3} \pder{x_I}
\end{align}
whereas the terms with derivatives with respect to $x_J$, $J \neq I$, are
\begin{align}
\label{eq: T comm A part 4}
-4 \; \zeta^{-4} \frac{f'(1/\zeta)^2}{(f(1/\zeta)-x_I)^2} \;
    \sum_{J \neq I} \frac{1}{x_J - x_I} \; \pder{x_J}
\textrm{ .}
\end{align}
We write the pure multiplication operator terms slightly suggestively
separating parts proportional to $(\delta_{x_J} - h_{x_J})$
\begin{align}
\nonumber
& \zeta^{-4} \frac{f'(1/\zeta)^2
    ( 8 \delta_{x_I} - 3 \kappa_I \delta_{x_I} )}{(f(1/\zeta)-x_I)^4}
+ 4 \sum_{J \neq I} \zeta^{-4}
    \frac{f'(1/\zeta)^2 \; h_{x_J}}{(f(1/\zeta)-x_I)^2 (x_J-x_I)^2} \\
\label{eq: T comm A part 5}
& \qquad - 4 \sum_{J \neq I} \zeta^{-4}
    \frac{f'(1/\zeta)^2 \, (\delta_{x_J}-h_{x_J})}
        {(f(1/\zeta)-x_I)^2 (f(1/\zeta)-x_J) (x_J-x_I)}
\textrm{ .}
\end{align}

We combine the contributions
(\ref{eq: T comm A part 1}), (\ref{eq: T comm A part 2}),
(\ref{eq: T comm A part 3}), (\ref{eq: T comm A part 4}) and 
(\ref{eq: T comm A part 5})
to yield a formula for the commutator
\begin{align*}
[\sT (\zeta) , \sA_I] \; = \; & \zeta^{-4}
    \frac{f'(1/\zeta)^2}{(f(1/\zeta)-x_I)^2} \Big( -2 \; \sA_I
    - 4 \sum_{J \neq I} \frac{\delta_{x_J}-h_{x_J}}
        {(f(1/\zeta)-x_J) (x_J-x_I)} \\
& \quad + \frac{6 -\kappa_I - 2 \kappa_I \delta_{x_I}}
        {f(1/\zeta)-x_I} \pder{x_I}
    + \frac{c + 8 \delta_{x_I} - 3 \kappa_I \delta_{x_I}}
        {(f(1/\zeta)-x_I)^2} \Big)
\textrm{ .}
\end{align*}
It is the first term that we wanted. The other terms vanish if
$\delta_{x_J} = h_{x_J}$ for $J \neq I$,
$\delta_{x_I} = \frac{6 - \kappa_I}{2 \kappa_I}$ and
$c = \frac{(6-\kappa_I) (3 \kappa_I - 8)}{2 \kappa_I}$
as claimed.
We once again point out that for a fixed central charge $c$ there
are two allowed values of $\kappa$, those dual to each other
via $\kappa^* = 16/\kappa$.
\QED


\section{Explicit expressions for $\sL_n$}
\label{app: explicit L}
The following table shows explicitly expressions for the 
operators $\sL_n$, $n \geq -2$.
One should interpret $f_0=1$ and $f_{-1} = 0$ whenever such factors
appear
\begin{eqnarray*}
\sL_n & = & - \sum_{l \leq -n} (1+n+l) f_{n+l}
    \pder{f_l}
\qquad \textrm{ for $n \geq 2$.} \\
\sL_1 & = & - \sum_{l \leq -3} (2+l) f_{1+l}
    \pder{f_l} + \sum_I \Big( \pder{x_I} \Big)
    + \sum_K \Big( \pder{y_K} \Big) \\
\sL_0 & = & - \sum_{l \leq -2} l f_{l} \pder{f_l}
    + \sum_I \Big( x_I \pder{x_I} + \delta_{x_I} \Big)
    + \sum_K \Big( y_K \pder{y_K} + \delta_{y_K} \Big) \\
\sL_{-1} & = & - \sum_{l \leq -2} \big( l f_{l-1} -
    \sum_{\substack{ m_1, m_2 \leq 0 \\ m_1+m_2 = l-1}} f_{m_1} f_{m_2}
    \big) \pder{f_l} \\
& & + \sum_I \Big( \big( x_I^2 - 3 f_{-2} \big) \pder{x_I}
        + (2 x_I) \delta_{x_I} \Big)
    + \sum_K \Big( \big( y_K^2 - 3 f_{-2} \big) \pder{y_K}
        + (2 y_K) \delta_{y_K} \Big)  \\
\sL_{-2} & = & - \frac{c}{2} f_{-2} - \sum_{l \leq -2} \big( (l-1) f_{l-2}
    - \sum_{\substack{ m_1, m_2, m_3 \leq 0 \\ m_1+m_2+m_3 = l-2}}
    f_{m_1} f_{m_2} f_{m_3} + 4 f_{-2} f_{l} \big) \pder{f_l} \\
& & + \sum_I \Big( \big( x_I^3 - 4 x_I f_{-2} -5 f_{-3} \big) \pder{x_I}
    + (3 x_I^2 - 4 f_{-2}) \delta_{x_I} \Big) \\
& & + \sum_K \Big( \big( y_K^3 - 4 y_K f_{-2} -5 f_{-3} \big) \pder{y_K}
    + (3 y_K^2 - 4 f_{-2}) \delta_{y_K} \Big) 
\textrm{ .}
\end{eqnarray*}



\def\cprime{$'$} \def\cprime{$'$} \def\cprime{$'$}

\end{document}